\documentclass[prd, reprint]{revtex4-1}

\usepackage[utf8]{inputenc}
\usepackage{graphicx}
\usepackage{verbatim}
\usepackage[T1]{fontenc}
\usepackage[english]{babel}
\usepackage{bm}
\usepackage{enumerate}
\usepackage{amsmath}
\usepackage{amssymb}
\usepackage{amsbsy}
\usepackage{amsthm}
\usepackage{bbold}
\usepackage{color}
\usepackage{subfigure}
\usepackage[shortcuts]{extdash}
\usepackage{upgreek}
\usepackage{mathtools}
\usepackage[colorlinks=true, citecolor=blue]{hyperref}
\usepackage[normalem]{ulem}
\usepackage[dvipsnames]{xcolor}
\usepackage{soul}

\DeclareMathOperator{\Tr}{Tr}
\newcommand{\qhat}{\hat{q}}

\newcommand{\dg}{\dagger}
\newcommand{\ev}[1]{\big<  #1 \big> }

\newcommand{\p}{\partial}
\newcommand{\one}{\mathbb{1}}

\newcommand{\vv}[1]{\mathbf{#1}}
\newcommand{\tf}[1]{\mathbf{#1}}

\begin{document}
\title{
Anisotropic momentum broadening in the 2+1D Glasma: \\ analytic weak field approximation and lattice simulations}
\author{A.~Ipp}
\email{ipp@hep.itp.tuwien.ac.at}
\author{D.~I.~Müller}
\email[Corresponding author: ]{dmueller@hep.itp.tuwien.ac.at}
\author{D.~Schuh}
\email{schuh@hep.itp.tuwien.ac.at}
\affiliation{Institute for Theoretical Physics, TU Wien, Austria}
\date{\today}

\begin{abstract}
    In heavy ion collisions, transverse momentum broadening quantifies the modification of a hard probe due to interactions with the quark-gluon plasma (QGP).
    We calculate momentum broadening in the Glasma, which is the highly non-isotropic precursor stage of the QGP right after the collision.
    We show that the Glasma leads to anisotropic momentum broadening: high energy partons accumulate more momentum along the beam axis than transverse to it.
    The physical origin of anisotropic broadening can be traced back to differences in the shapes of chromo-electric and chromo-magnetic flux tubes in the Glasma.
    We provide semi-analytic results for momentum broadening in the dilute Glasma and numerical results from real-time lattice simulations of the non-perturbative dense Glasma.
\end{abstract}

\maketitle

\section{Introduction}
Understanding the properties of QCD matter in heavy-ion collisions provides many challenges and requires a combination of multiple approaches and tools for the different stages involved. An important probe that can be sensitive to all stages involved is high-energetic jets. They originate from initial hard scatterings and probe the emerging and evolving medium throughout its evolution before they hit the detectors in sprays of particles.
The reduction of the energy through medium effects is known as jet quenching \cite{Mehtar-Tani:2013pia, Qin:2015srf}.
Experimental signatures include suppression of hadron spectra at high transverse momentum $p_{T}$, modification of back-to-back correlations, and modification of reconstructed jets in nucleus-nucleus collisions as compared to proton-proton collisions.
These effects have been extensively studied in the quark-gluon plasma stage using combined models, where the perturbative QCD effects stem at leading order from collisional energy loss and medium-induced gluon radiation \cite{Baier:1996kr,Schenke:2009ik,Kurkela:2014tla,Niemi:2015qia,Andrews:2018jcm}.
Recently, it has been shown that the effect of jet quenching can be very sensitive to pre-hydrodynamic time-scales if the nuclear modification factor $R_{AA}$ and high-$p_{T}$ elliptic flow $v_{2}$ have to be matched simultaneously \cite{Andres:2019eus}.
Therefore, a complete treatment of jet quenching should also consider modifications from the later hadron gas stage \cite{Dorau:2019ozd} as well as from earlier stages before hydrodynamics becomes applicable.

The Glasma \cite{Lappi:2006fp, Gelis:2012ri} is a pre-hydrodynamic stage based on the color glass condensate (CGC) effective theory. It differs from the later quark-gluon plasma phase by non-equilibrated, highly anisotropic flux tubes with longitudinal chromo-magnetic and chromo-electric fields.

A first qualitative estimate of jet quenching from the Glasma has been given by considering synchrotron-like gluon emission in the coherent color fields of the Glasma \cite{Aurenche:2012qk}. Such contributions turned out to be smaller than the radiative energy loss in the QGP phase, but this result relies on a small angle approximation which strictly only holds for high-$p_{T}$ hadrons. In an energy regime $\omega\lesssim5$ GeV, which is still important for the total energy loss, the contributions from the Glasma phase may no longer be negligible. 
Momentum broadening and energy loss due to collisions have been solved on a lattice using the Wong-Yang-Mills equations with hard collisions among the partons \cite{Schenke:2008gg}. The estimate provided for a jet traversing a classical Yang-Mills field indicates a significant fractional energy loss of 10\% -- 20\% over the first $\tau\simeq 1$ fm/c. One should take into account though that \cite{Schenke:2008gg} did not use typical initial conditions for the Glasma and did not account for the longitudinal expansion of the system.
A proper time expansion has been taken into account in a recent calculation for the leading analytic contribution for heavy quarks traversing the Glasma \cite{Carrington:2020sww}. However, the truncation used is valid only for very small times and gauge invariance is not apparent in the chosen set-up.

In this work, we study the contribution of momentum broadening of a parton in the evolving Glasma in a systematic and gauge-invariant way. We compare analytic calculations with real-time lattice simulations and verify that they agree in the dilute limit. We observe anisotropic momentum broadening with larger momentum growth along the beam axis than transverse to it. This effect is also observed in our lattice simulations for the dense case and can be traced back to initial differences in the chromo-electric and chromo-magnetic field correlators of the Glasma.

The source of the anisotropic transverse momentum broadening that we find in the Glasma differs from various origins and mechanisms that have been proposed for the quark-gluon plasma: in \cite{Romatschke:2006bb} a calculation based on kinetic theory shows that the anisotropic momentum distribution in a longitudinally expanding quark-gluon plasma leads to larger jet broadening along the beam axis. Instabilities in non-Abelian plasmas \cite{Dumitru_2008} and turbulent color fields \cite{PhysRevLett.99.042301} can also give rise to anisotropic broadening. Similar effects have been found in anisotropic supersymmetric Yang-Mills plasmas based on the gauge/gravity duality \cite{2012JHEP...07..031G, 2012JHEP...08..041C, Rebhan:2012bw}. Furthermore, the flow of the medium can deform the energy and momentum distribution of a jet \cite{Armesto:2004pt}. In contrast, the physical origin of anisotropic momentum broadening investigated in this paper is related to particular properties of Glasma flux tubes. The model for the Glasma we consider is homogeneous in the transverse plane. Hence, no net flow is possible which allows us to rule out effects due to the flow of the medium. Moreover, we show that the anisotropy is unrelated to the longitudinal expansion of the Glasma. 
The anisotropy is directly linked to differences in the average spatial shapes of chromo-electric and \-/magnetic Glasma flux tubes (more precisely, their respective spatial autocorrelation functions) which exert different Lorentz forces on moving color charges.

This paper is organized as follows. 
In Section \ref{sec:formalism} we lay out the general formalism to calculate momentum broadening for a test particle traversing through a non-Abelian background field. The formulae that we develop are manifestly gauge invariant. In Section \ref{sec:glasma} we apply this formalism to the Glasma background field in a weak-field approximation and in the lattice formalism. In Section \ref{sec:results} we finally present numerical checks of weak-field results with lattice simulations in the dilute limit and provide results for the dense case.

\section{Basic formalism \label{sec:formalism}}
In this section we present the basic formalism which we use to compute transverse momentum broadening of highly relativistic test particles as they traverse the Glasma.
The methods we use are largely based on the definition of momentum broadening via a lightlike Wilson loop \cite{Liu:2006ug, CasalderreySolana:2007qw, Panero:2013pla} and, equivalently, the dynamics of classical colored particles \cite{Majumder:2009cf, Carrington:2016mhd, Mrowczynski:2017kso, Ruggieri:2018rzi, Sun:2019fud}. We review these approaches and derive a gauge invariant expression for transverse momentum broadening of an ultrarelativistic colored particle.

\begin{figure}
    \centering
    \includegraphics[scale=0.2]{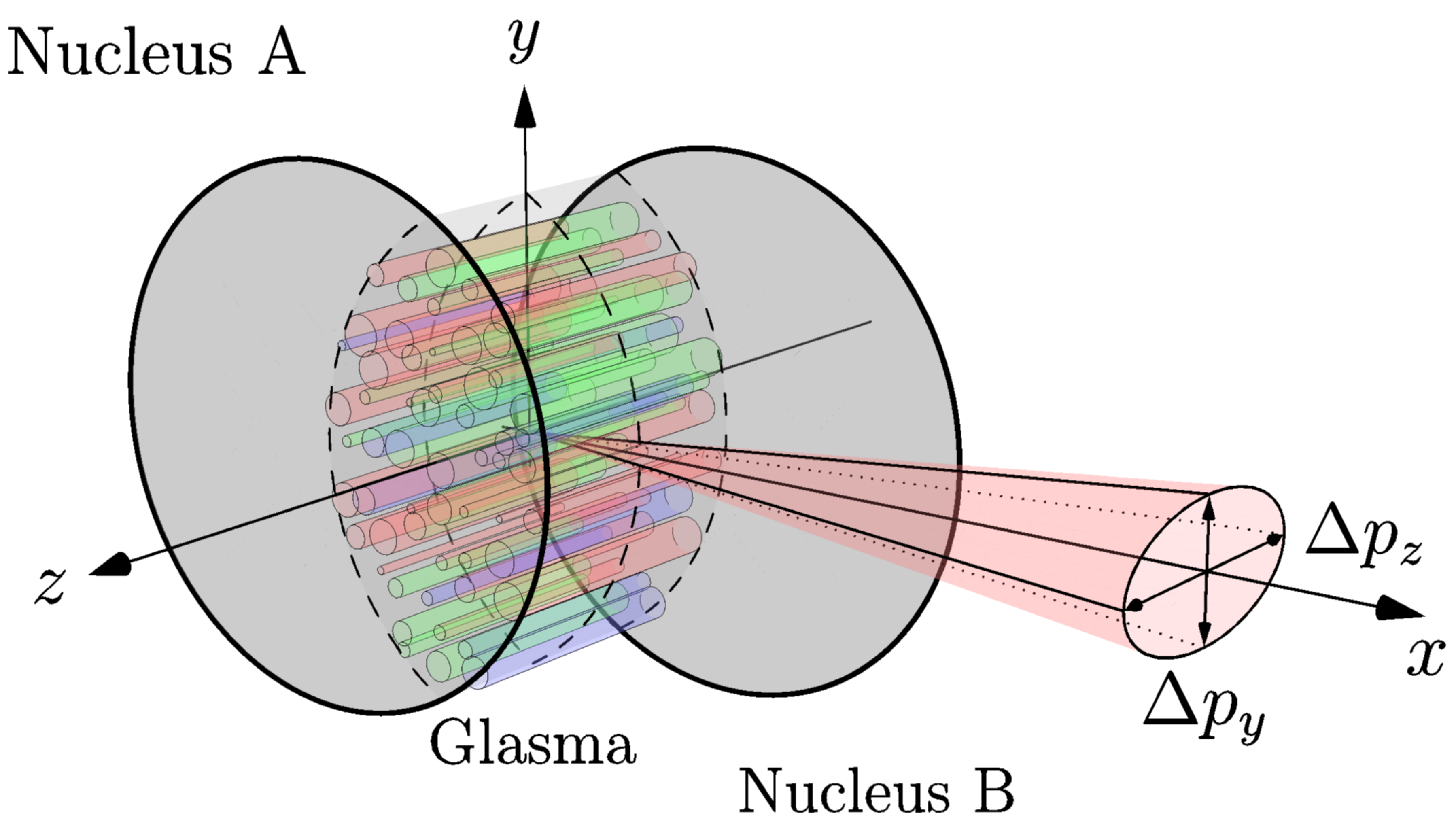}
    \caption{Momentum broadening of fast partons from interaction with the Glasma. In this schematic we show a high energy particle created in the collision of two nuclei escaping towards the detector along the $x$ axis with initial momentum $p_x$. The collision of the two nuclei produce the Glasma: a purely gluonic, highly occupied anisotropic precursor state to the quark-gluon plasma. As the particle moves through and interacts with the Glasma, it accumulates transverse momentum (in the sense of orthogonal to $x$, i.e.~$p_y$ and $p_z$). This leads to momentum broadening $\Delta p_z$ along the beam axis $z$ and broadening $\Delta p_y$ within the transverse plane (spanned by $x$ and $y$).}
    \label{fig:momentum_broadening}
\end{figure}

The geometry of the problem is shown in fig.~\ref{fig:momentum_broadening}. Two high energy nuclei move along the beam axis $z$ and collide at $z=0$ at time $t=0$. During the collision, hard partons from the colliding nuclei can scatter and produce high energy jets. In fig~\ref{fig:momentum_broadening}, a single particle from such a scattering (either a quark or a gluon) is shown, which moves along the $x$-axis. We assume that this particle has a high initial momentum $p_x$ and zero momentum in the two orthogonal directions, $p_y=p_z=0$. Around the same time, the Glasma is formed from interactions among the soft partons of the nuclei. The Glasma is a purely gluonic, highly occupied state and can be described as a classical color field. Initially, the Glasma consists of chromo-electric and chromo-magnetic flux tubes aligned with the beam axis $z$. In fig.~\ref{fig:momentum_broadening} these flux tubes are depicted as colorful cylinders. 
As the system evolves and the particle traverses the Glasma, it interacts with the color fields of the Glasma and accumulates momentum along its path due to the non-Abelian Lorentz force exerted on the particle. For sufficiently high initial momentum $p_x$, the particles' trajectory is almost unaffected by this interaction, but over time the particle accumulates transverse momentum $p_y$ and $p_z$.
This leads to a momentum broadening effect depicted in fig.~\ref{fig:momentum_broadening} as a red cone around the $x$ axis.

A note of caution is necessary regarding the terms \textit{longitudinal} and \textit{transverse}. In the context of jet momentum broadening, the terms transverse and longitudinal are defined relative to the jet trajectory. Using this nomenclature the $x$ axis in the situation shown in fig.~\ref{fig:momentum_broadening} would be the longitudinal axis, while $y$ and $z$ are transverse coordinates. This should not be confused with the longitudinal and transverse directions relative to the beam axis as used in the context of Glasma calculations. In the Glasma the longitudinal axis is identified with the beam axis $z$, while $x$ and $y$ span the transverse plane. For example, transverse momenta of gluons in the Glasma refer to momenta within the $xy$-plane. Transverse momentum broadening of a jet moving along $x$ refers to broadening of momenta in the $yz$-plane.

\subsection{Classical approach to momentum broadening}

We start by sketching the basic derivation of momentum broadening in a more general context before focusing on non-Abelian background fields and eventually the Glasma. Consider an ultrarelativistic test particle traveling through a medium, for example a high momentum quark through the Glasma.
It follows a straight, lightlike trajectory parametrized by 
\begin{equation}
x^\mu(t) = u^\mu t + x^\mu_0,
\end{equation}
where $t$ is the time coordinate in the rest frame of the medium, the tangent vector $u^\mu$ is constant and lightlike, i.e.\ $u^2 = u_\mu u^\mu = 0$, and $x^\mu_0$ defines the starting point of the particle at $t = 0$.

Without loss of generality we choose $u^\mu = (1, 1, 0, 0)^\mu$ and $x_0^\mu = 0$. In this context, $x^1$ is referred to as the longitudinal coordinate, while $x^2$ and $x^3$ are referred to as transverse coordinates. The longitudinal part of the trajectory of the particle reads $x^1(t) = t$, where we have set the speed of light $c$ to one, i.e.\ $c = 1$.
While the particle is moving through the medium, forces acting on the particle would deflect it from its straight trajectory.
Assuming that the particle is highly energetic, the force has almost no effect on the trajectory if considered over short periods of time. However, we can still obtain an estimate on the momentum that the particle accumulates along its path.

The classical equation of motion of the particle in the frame of the medium
\begin{equation}
\frac{d p^\mu}{dt} = \mathcal{F}^\mu(x(t))
\end{equation}
is formally solved by
\begin{equation}
p^\mu(t) = p^\mu(0) + \intop_0^t dt' \mathcal{F}^\mu(x(t')),
\end{equation}
where $p^\mu = (p^0, p_\parallel, \vv p_\perp)$ is the particle's four-momentum and $\mathcal{F}^\mu(x(t))$ is an external force due to interactions with the medium. Here $p_\parallel = p_x$ is the longitudinal momentum and $\vv p_\perp = (p_y, p_z)$ is the transverse momentum of the particle with respect to its trajectory.
Assuming that the particle starts without transverse momentum, i.e.~$\vv p_\perp(0) = \vv 0$, we find
\begin{equation}
p_i(t) = \intop_0^t dt' \mathcal{F}_i (x(t')),
\end{equation}
where $i \in \{2, 3\}$ is a transverse coordinate index. The force $\mathcal{F}_i$ exerted on the particle, when computed from the Glasma, can be considered a random variable with zero mean
\begin{equation} \label{eq:langevin_force_1}
\ev{\mathcal{F}^\mu} = 0,
\end{equation}
but non-zero two-point function
\begin{equation} \label{eq:langevin_force_2}
\ev{\mathcal{F}^\mu(x) \mathcal{F}^\nu(y)} \neq 0.
\end{equation}
Here, $\ev{\dots}$ refers to the average over all realizations of the medium (e.g.~the Glasma in our case). We stress that eqs.~\eqref{eq:langevin_force_1} and \eqref{eq:langevin_force_2} are consequences of the Glasma picture (see section \ref{sec:glasma}) and should not be seen as additional assumptions. 
As a result, the averages of the transverse components $\ev{p_i(t)}$ vanish, but their variances $\ev{p^2_i(t)}$ are given by
\begin{equation} \label{eq:force_correlator_components}
\ev{p^2_{i}(t)} = \intop_0^t dt' \intop_0^t dt''  \ev{\mathcal{F}_{i}(x(t'))  \mathcal{F}_{i}(x(t''))},
\end{equation}
where no sum over the index $i \in \{2, 3\}$ is implied. The total transverse momentum is then given by
\begin{equation} \label{eq:acc_momentum_sum}
\ev{\vv p^2_\perp(t)} = \sum_{i \in \{ 2,3 \}} \ev{p^2_i(t)}.
\end{equation}
The (instantaneous) jet broadening parameter $\qhat(t)$ is defined as the time derivative of the accumulated transverse momentum (see e.g.~\cite{Carrington:2016mhd})
\begin{equation}
\qhat(t) = \frac{d}{dt} \ev{\vv p^2_\perp(t)}.
\end{equation}
More generally, one can define a jet broadening parameter with respect to the transverse axis $x^i$ by taking the time derivative of a single momentum component
\begin{equation}
\qhat_i(t) = \frac{d}{dt} \ev{p^2_{i}(t)}.
\end{equation}
Analogous to eq.~\eqref{eq:acc_momentum_sum} it holds that
\begin{equation}
\qhat(t) = \sum_{i\in \{2,3 \}} \qhat_i(t).
\end{equation}
As we will see in section \ref{sec:results}, it is reasonable to distinguish between the momentum broadening in the two transverse directions due to the special field configuration of the Glasma present immediately after the collision.

\subsection{Momentum broadening in an Abelian background field} \label{sec:tmb_abelian}

\begin{figure}
    \centering
    \includegraphics{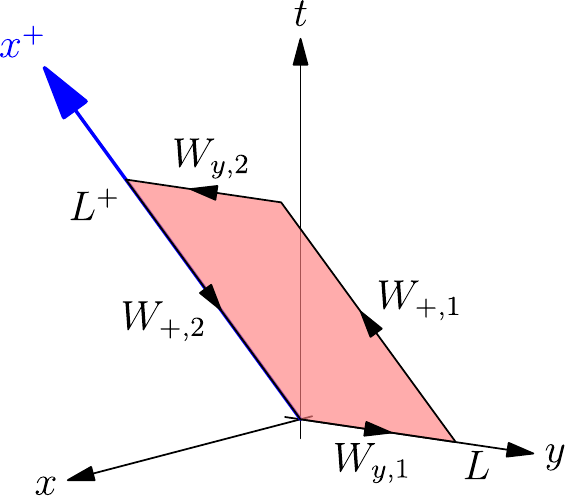}
    \caption{The rectangular Wilson loop associated with momentum broadening in the $y$ direction with transverse extent $L$ along $y$ and a path length of $L^+$ along $x^+$. In our derivation the Wilson loop is collapsed to a single line along $x^+$ by performing a Taylor expansion in the transverse extent $L$. }
    \label{fig:wilson_loop}
\end{figure}

Before tackling the problem of non-Abelian fields, we derive $\ev{p^2_i(t)}$ for an electrically charged particle in an Abelian background field $A_\mu$. Using the same tangent vector as before, $u^\mu = (1, 1, 0, 0)^\mu$, the Abelian Lorentz force acting on a particle with charge $q$ is given by
\begin{equation}
\mathcal{F}_i(x(t)) = q u^\nu F_{i\nu}(x(t)),
\end{equation}
where $F_{\mu\nu} = \p_\mu A_\nu - \p_\nu A_\mu$ is the Abelian field strength tensor associated with the background field $A_\mu$. Introducing the light cone coordinates $x^+$ and $x^-$ relative to the particle trajectory via
\begin{equation} \label{eq:lc1}
x^\pm = \frac{x^0 \pm x^1}{\sqrt{2}},
\end{equation}
we can write
\begin{equation}
\mathcal{F}_i(x(t)) = \sqrt{2} q F_{i+}(x(t)).
\end{equation}
Inserting this into eq.~\eqref{eq:force_correlator_components} yields
\begin{equation} \label{eq:psquared_Abelian_particle}
\ev{p^2_{i}(t)} = 2 q^2 \intop_0^t dt' \intop_0^t dt''  \ev{
F_{i+}(x(t')) F_{i+}(x(t''))
},
\end{equation}
where, as before, no sum over the transverse index $i$ is implied.

The result, eq.~\eqref{eq:psquared_Abelian_particle}, obtained using the classical equations of motion for a charged particle, can also be found using the dipole approximation \cite{Liu:2006ug} via the expansion of a Wilson loop with lightlike edges aligned with the particle trajectory. A sketch of this Wilson loop is shown in fig.~\ref{fig:wilson_loop}. Consider the rectangular Wilson loop $W_{y+}$ spanned across one of the two transverse coordinates $y$ (from $y = 0$ to $y = L$) and the lightlike axis $x^+$ (from $x^+ = 0$ to $x^+ = L^+$) given by
\begin{align} \label{eq:abelian_wilson_line}
W_{y+} &= \exp{\bigg( - i q \oint dx^\mu A_\mu \bigg)} \nonumber \\
&=W_{+, 2} W_{y, 2} W_{+, 1} W_{y, 1},
\end{align}
where the transverse Wilson lines are
\begin{align}
W_{y, 1} &= \exp{\bigg(-i q \intop_0^{L} dy A_y(0, x^-, y, z)\bigg)}, \\
W_{y, 2} &= \exp{\bigg(+i q \intop^L_{0} dy A_y(L^+, x^-, y, z)\bigg)},
\end{align}
and the lightlike Wilson lines are given by
\begin{align}
W_{+, 1} &= \exp{\bigg(-i q \intop_0^{L^+} dx^+ A_+(x^+, x^-, L, z)\bigg)}, \\
W_{+, 2} &= \exp{\bigg(+i q \intop^{L^+}_0 dx^+ A_+(x^+, x^-, 0, z)\bigg)}.
\end{align}

In the dipole approximation this Wilson loop can be related to the squared momentum $\ev{p_y^2}$ via
\begin{equation} \label{eq:abelian_dipole}
\ev{\operatorname{Re} W_{y+}} = \exp{\bigg( - \frac{L^2 \ev{p^2_y}}{2} \bigg)}
\end{equation}
for small transverse extensions of the loop $L \ll L^+$. To show this equivalence, we use Stokes' theorem to rewrite the integral around the closed, rectangular path as an area integral over the field strength tensor
\begin{align}
W_{y+} &= \exp{\bigg( - i q \intop_0^{L^+} dx^+ \intop_0^{L} dy \, \left(\p_y A_+ - \p_+ A_y\right)\bigg)} \nonumber \\
&= \exp{\bigg( - i q \intop_0^{L^+} dx^+ \intop_0^{L} dy \, F_{y+} \bigg)}.
\end{align}
\begin{widetext}
Taking the real part of $W_{y+}$ and expanding for small transverse distances $L \ll L^+$ yields
\begin{align}
\operatorname{Re} W_{y+} &=1 - \frac{q^2 L^2}{2} \intop_0^{L^+} dx'^+ \intop_0^{L^+} dx''^+ F_{y+}(x'^+, x^-, y, z) F_{y+}(x''^+, x^-, y, z) + \mathcal{O}(L^4).
\end{align}
Performing the medium average and rewriting the integrals along $x^+$ as an integration along the lightlike trajectory of the particle $x(t)$, we find
\begin{equation}
\ev{\operatorname{Re} W_{y+}} = 1 - q^2 L^2 \intop_0^t dt' \intop_0^t dt'' \ev{F_{y+}(x(t')) F_{y+}(x(t''))} + \mathcal{O}(L^4).
\end{equation}
\end{widetext}
On the other hand, we can expand eq.~\eqref{eq:abelian_dipole} up to second order in $L$:
\begin{equation} \label{eq:collapsed_wilson_loop}
\ev{\operatorname{Re} W_{y+}} = 1 - \frac{L^2 \ev{p_y^2}}{2} + \mathcal{O}(L^4).
\end{equation}
Comparing coefficients yields
\begin{equation}
\ev{p_y^2(t)} = 2 q^2 \intop_0^t dt' \intop_0^t dt'' \ev{F_{y+}(x(t')) F_{y+}(x(t''))}.
\end{equation}
This derivation can be repeated for the Wilson loop $W_{i+}$, where the transverse extent is aligned with the $x^i$ axis with $i\in \{y, z\}$. The general result reads
\begin{equation} \label{eq:psquared_correlator_abelian}
\ev{p_i^2(t)} = 2 q^2 \intop_0^t dt' \intop_0^t dt'' \ev{F_{i+}(x(t')) F_{i+}(x(t''))},
\end{equation}
where no sum over $i$ is implied.
We therefore recovered eq.~\eqref{eq:psquared_Abelian_particle}, which we previously derived using classical particle dynamics.

More generally, if we align the transverse extension of the Wilson loop along some other transverse direction, we obtain the expression
\begin{equation} \label{eq:abelian_wilson_loop}
\ev{\operatorname{Re} W_{i+}} = \exp{\bigg( - \frac{L^2 \ev{p^2_i}}{2} \bigg)}
\end{equation}
for $L \ll L^+$, where $L$ denotes the length of the loop along $x^i$. In order to directly compute $\ev{p_i^2}$ we expand the Wilson loop $W_{i+}$ as a function of $L$
\begin{equation}
W_{i+} = 1 + L W^{(1)}_{i+} + \frac{L^2}{2} W^{(2)}_{i+} + \mathcal{O}(L^3),
\end{equation}
and take the expectation value of the quadratic term:
\begin{equation} \label{eq:psquared_quadratic_coeff_abelian}
\ev{p^2_i(t)} = - \ev{ \operatorname{Re} \left[ W^{(2)}_{i+} \right]}.
\end{equation}

\subsection{Momentum broadening in a non-Abelian background field}

\subsubsection{Derivation from the Wilson loop}

The results of the previous sections can be generalized to non-Abelian fields and particles with color charge. In Yang-Mills theory with gauge group $\mathrm{SU}(N_c)$, Wilson loops are elements of $\mathrm{SU}(N_c)$ and the trace over a Wilson loop depends on the chosen representation $R$ of $\mathrm{SU}(N_c)$. A straightforward generalization of eq.~\eqref{eq:abelian_wilson_loop} is given by
\begin{equation}
\frac{1}{D_R} \ev{\operatorname{Re} \Tr \left[ W_{i+} \right]_R } = \exp{\bigg( -\frac{L^2}{2} \ev{p^2_i}_R \bigg)},
\end{equation}
where $D_R$ is the dimension of the representation and $\Tr \left[ \dots \right]_R$ denotes the trace operation in the given representation. The choice of representation  of $\mathrm{SU}(N_c)$ decides the species of particle that we consider. For quarks we choose the fundamental representation $F$ and for gluons we choose the adjoint representation $A$. We add a subscript $R$ in $\ev{p^2_i}_R$ to denote that the accumulated squared momentum depends on the representation. 
As before, $\ev{p^2_i}_R$ is given by the second coefficient of the expansion in $L$:
\begin{equation} \label{eq:psquared_quadratic_nonabelian}
\ev{p^2_i(t)}_R = - \frac{1}{D_R} \ev{ \operatorname{Re}  \Tr \left[ W^{(2)}_{i+} \right]_R },
\end{equation}
where
\begin{equation} \label{eq:wilson_loop_expansion_nonabelian}
W_{i+} = \one + L W^{(1)}_{i+} + \frac{L^2}{2} W^{(2)}_{i+} + \mathcal{O}(L^3).
\end{equation}
In fact, it is not necessary to actually compute the quadratic coefficient in order to obtain $\ev{p^2_i(t)}$. Enforcing the unitary constraint
\begin{equation}
W_{i+} W^\dg_{i+} = \one 
\end{equation}
for the expansion eq.~\eqref{eq:wilson_loop_expansion_nonabelian}, yields two constraints for the Taylor coefficients
\begin{align}
\left[ W^{(1)}_{i+} \right]^\dg &= -  W^{(1)}_{i+}, \\
\frac{1}{2} \left( W^{(2)}_{i+} +  W^{(2)\dg}_{i+} \right) &= \left( W^{(1)}_{i+} \right)^2.
\end{align}
Using the second relation we can write
\begin{align}
\ev{p^2_i(t)}_R &= - \frac{1}{D_R} \ev{ \operatorname{Re}  \Tr \left[ W^{(2)}_{i+} \right]_R } \nonumber \\
&= - \frac{1}{D_R} \ev{\Tr \left[ \left( W^{(1)}_{i+} \right)^2 \right]_R}.  \label{eq:psquared_linear_nonableian}
\end{align}
The accumulated momenta $\ev{p^2_i(t)}$ are therefore given by the linear coefficient of the Wilson loop, which drastically simplifies the calculation.

A detailed derivation of the linear coefficient $W_{i+}$ can be found in appendix \ref{sec:app_wilson}. The final result for the squared accumulated momentum (see eq.~\eqref{eq:psquared_correlator_nonabelian_app}) reads
\begin{align} \label{eq:psquared_correlator_nonabelian}
&\ev{p^2_i(t)}_R = \nonumber \\
&\quad\frac{2 g^2}{D_R} \intop_0^t dt' \intop_0^t dt'' \ev{\Tr \bigg[  \widetilde{F}_{i+}(x(t')) \widetilde{F}_{i+}(x(t''))\bigg]_R },
\end{align}
where we have defined the parallel transported field strength tensor
\begin{align} \label{eq:parallel_transported_F}
\widetilde{F}_{i+}(x) &= W_+(0, x^+) F_{i+}(x) W_+(x^+, 0),
\end{align}
and the lightlike Wilson line $W_+(b, a)$ connecting $x^+=a$ to $x^+=b$
\begin{equation}
W_+(b, a) = \mathcal{P} \exp{\bigg( - i g \intop_a^{b} dx^+ A_+(x^+) \bigg)}.
\end{equation}
Here, $\mathcal{P}$ denotes path ordering along $x^+$. In the above expression, the gauge field $A_+$ is evaluated along the lightlike particle trajectory.  

Our main result eq.~\eqref{eq:psquared_correlator_nonabelian} is the non-Abelian generalization of eq.~\eqref{eq:psquared_correlator_abelian}. It is evident that this expression for $\ev{p^2_i(t)}_R$ is gauge invariant: the Wilson lines in eq.~\eqref{eq:parallel_transported_F} ensure that under a general gauge transformation $\Omega(x)$, the parallel transported field strength tensor transforms according to
\begin{equation}
\widetilde{F}_{i+}(x) \rightarrow \Omega(x^+=0) \widetilde{F}_{i+}(x) \Omega^\dg(x^+=0).
\end{equation}
The trace over the square of the field strength tensor in ~eq.~\eqref{eq:psquared_correlator_nonabelian} is then guaranteed to be gauge invariant.

\subsubsection{Derivation using classical colored particle dynamics}

Having derived eq.~\eqref{eq:psquared_correlator_nonabelian} using the Wilson loop, we now show that the same result can be found using classical colored particle dynamics. The equivalent derivation is based on colored particles affected by a non-Abelian Lorentz force as given by Wong's equations
\begin{align}
\frac{dp^\mu}{dt} &= g Q^a(t) \mathcal{F}^{a,\mu} (x(t)), \\
\frac{dQ^a(t)}{dt} &= g \frac{dx^\mu(t)}{dt} f^{abc}  A^b_\mu(x(t)) Q^c(t),
\end{align}
where the non-Abelian force is given by
\begin{equation}
\mathcal{F}^{a,\mu}(x(t)) = \frac{dx_\nu (t)}{dt} F^{a,\mu\nu}(x(t)).
\end{equation}
Here, $Q^a(t)$ are the components of the time-dependent color charge of the particle and $f^{abc}$ are the anti-symmetric structure constants of $\mathrm{SU}(N_c)$. In order to  rewrite everything in terms of group and algebra elements of $\mathrm{SU}(N_c)$, we use
\begin{align}
Q &= Q^a \tf T^a, \\
F^{\mu\nu} &=  F^{a,\mu\nu} \tf T^a,
\end{align}
where $\tf T^a$ are the generators of $\mathrm{SU}(N_c)$, and the relation
\begin{equation}
\Tr \left[\tf  T^a \tf  T^b \right]_R = T_R \delta^{ab},
\end{equation}
where $R \in \{ F, A \}$ and $T_F = 1/2$ and $T_A = N_c$ in the fundamental and adjoint representations respectively. We can write the equations of motion as
\begin{align}
\frac{dp^\mu}{dt} &= \frac{g}{T_R} \frac{dx_\nu (t)}{dt} \Tr \left[ Q(t) F^{\mu\nu}(x(t)) \right]_R, \\
\frac{dQ(t)}{dt} &= - i g \frac{dx^\mu(t)}{dt} \left[A_\mu(t), Q(t) \right].
\end{align}
The formal solution to the second equation is given by
\begin{equation}
Q(t) = U(t, 0) Q(0) U(0, t),
\end{equation}
where
\begin{equation}
U(t, 0) = \mathcal{P} \exp{\left( - i g \intop^t_0 dt' \frac{dx^\mu(t')}{dt'} A_\mu(x(t')) \right)}.
\end{equation}
If we choose a lightlike trajectory, i.e. $x^\mu(t) = u^\mu t$ with $u^\mu = (1, 1, 0, 0)^\mu$, we recover the lightlike Wilson line
\begin{align}
U(t, 0) &= W_+(x^+(t), 0) \nonumber \\
&= \mathcal{P} \exp{ \bigg( - i g \intop_0^{x^+(t)} dx'^+ A_+(x'^+) \bigg)},
\end{align}
where $x^+(t) = \sqrt{2} t$. For readability, we suppress other coordinates on which $A_\mu$ depends. The equations of motion for the transverse momenta read
\begin{equation}
\frac{dp^i}{dt} = \frac{\sqrt{2} g}{T_R} \Tr \left[U(t,0) Q(0) U(0, t) F^i{}_+(x(t))\right]_R.
\end{equation}
Repeating the same steps as before, we find that the accumulated squared transverse momentum reads
\begin{widetext}
\begin{align}
\ev{p^2_i(t)} = \frac{2 g^2}{T^2_R} Q^a(0) Q^b(0) \intop_0^t dt' \intop_0^t dt'' \bigg< \Tr \left[ \tf T^a U(0, t') F_{i+}(x(t')) U(t', 0)\right]_R \Tr \left[ \tf T^b U(0, t'') F_{i+}(x(t'')) U(t'', 0) \right]_R \bigg>,
\end{align}
where $Q^a(0)$ are the components of the initial color charge of the particle and no sum over $i$ is implied. The next step, following \cite{Litim:2001db, Majumder:2009cf, Carrington:2016mhd}, is to perform an average over $Q^a(0)$ using
\begin{equation}
\int dQ Q^a Q^b = T_R \delta^{ab},
\end{equation}
and (as done in \cite{Carrington:2016mhd}) divide by the dimension of the representation $D_R$. This yields
\begin{align} \label{eq:psquared_nonabelian_cpd1}
\ev{p^2_i(t)}_R = \frac{2 g^2}{T_R D_R} \intop_0^t dt' \intop_0^t dt'' \bigg< \Tr \left[ \tf T^a U(0, t') F_{i+}(x(t')) U(t', 0)\right]_R  \Tr \left[ \tf T^a U(0, t'') F_{i+}(x(t'')) U(t'', 0) \right]_R \bigg>.
\end{align}
\end{widetext}
Using the relation
\begin{equation}
\Tr \left[ \tf T^a X \right]_R \Tr \left[ \tf T^a Y \right]_R = T_R \Tr \left[ X Y\right]_R,
\end{equation}
where $X$ and $Y$ are arbitrary elements of the Lie algebra of $\mathrm{SU}(N_c)$, we can further simplify eq.~\eqref{eq:psquared_nonabelian_cpd1}. The result reads
\begin{align} \label{eq:psquared_nonabelian_cpd2}
\ev{p^2_i(t)}_R &= \frac{2 g^2}{D_R} \intop_0^t dt' \intop_0^t dt'' \big< \Tr\big[  U(0, t') F_{i+}(x(t'))  \nonumber \\
& \qquad \times U(t', 0) U(0, t'') F_{i+}(x(t'')) U(t'', 0) \big]_R \big>,
\end{align}
If we define the parallel transported field strength tensor as in eq.~\eqref{eq:parallel_transported_F}, we recover the same result as eq.~\eqref{eq:psquared_correlator_nonabelian}:
\begin{align} \label{eq:psquared_correlator_nonabelian_2}
&\ev{p^2_i(t)}_R = \nonumber \\
& \quad \frac{2 g^2}{D_R} \intop_0^t dt' \intop_0^t dt'' \ev{\Tr \bigg[  \widetilde{F}_{i+}(x(t')) \widetilde{F}_{i+}(x(t''))\bigg]_R }.
\end{align}

\subsection{Casimir scaling of accumulated momenta \label{sec:scaling}}

Having derived an expression for $\ev{p^2_i}_R$ given a representation $R$ of $\mathrm{SU}(N_c)$, it is easy to see that the accumulated transverse momentum scales with the Casimir $C_R$ of the representation. Using eq.~\eqref{eq:psquared_linear_nonableian} and the fact that $W^{(1)}_{i+}$ is anti-hermitian and traceless, we can write
\begin{equation}
W^{(1)}_{i+} = i \tf T^a W_i^a,
\end{equation}
and consequently
\begin{equation}
\ev{p^2_i}_R = \frac{T_R}{D_R} \ev{W^a_i W^a_i}.
\end{equation}
If we look at the ratio of the accumulated momentum of a gluon and a quark, we find the well-known result
\begin{equation}
\frac{\ev{p^2_i}_A}{\ev{p^2_i}_F} = \frac{T_A D_F}{T_F D_A} = \frac{C_A}{C_F} =\frac{2 N_c^2}{N_c^2 - 1}.
\end{equation}
For $N_c = 3$, this factor evaluates to $9/4$.

\section{Momentum broadening in the Glasma} \label{sec:glasma}
In this section we apply eq.\ \eqref{eq:psquared_correlator_nonabelian_2} to the boost-invariant Glasma.  
As noted in the beginning of section \ref{sec:formalism}, care has to be taken with nomenclature: in the Glasma, transverse coordinates are usually $x=x^1$ and $y=x^2$, whereas $z=x^3$ is the longitudinal coordinate along the beam axis. In contrast, the transverse momentum $\ev{p_\perp^2}$ refers to the momentum perpendicular to the particle trajectory, which we take to be aligned with one of the transverse directions in the Glasma.
Here, we choose the particle to move along the $x$ axis.
Additionally, we use a different definition for light cone coordinates. They are given by
\begin{equation}
x^\pm = \frac{x^0 \pm x^3}{\sqrt{2}},
\end{equation}
where, compared to eq.~\eqref{eq:lc1}, we use $x^3$ in place of $x^1$. Note that $x^3 = z$ is the beam axis.

We start with a quick review of the boost-invariant color glass condensate framework \cite{Iancu:2002xk, Lappi:2006fp, Gelis:2010nm, Gelis:2016rnt} and the Glasma. In the CGC, the hard partons of a nucleus moving at the speed of light along the positive $z$ axis (referred to as nucleus ``A'') are described by a color charge density $\rho_A$, which is highly localized around $x^- = 0$. The color current of such a nucleus is given by (under the assumption of covariant gauge, i.e.\ $\p_\mu A^\mu = 0$)
\begin{equation}
J_A^\mu = \delta^{\mu+} \rho_A(x^-, \vv x),
\end{equation}
where $\vv x = (x, y)$ is the transverse coordinate vector. Analogously, the color current of nucleus ``B'', which moves in the opposite direction, is given by
\begin{equation}
J_B^\mu = \delta^{\mu-} \rho_B(x^+, \vv x).
\end{equation}
Associated with these currents $J^\mu_A$ and $J^\mu_B$ are the gauge fields $A^\mu_A$ and $A^\mu_B$, which represent the soft gluons of the nuclei.
In the appropriate light cone gauges $A^+ = 0$ for nucleus ``A'' and $A^- = 0$ for ``B'', the gauge fields become purely transverse and are given by
\begin{equation}
A^{(A,B)}_i(x^\mp, \vv x) = \frac{1}{ig} V_{(A,B)}(x^\mp, \vv x) \p^i V^\dg_{(A,B)}(x^\mp, \vv x).
\end{equation}
The lightlike Wilson lines along $x^\pm$ are given by
\begin{equation} \label{eq:cgc_wilson}
V^\dg_{(A,B)}(x^\mp, \vv x) = \mathcal{P} \exp{ \bigg( i g \intop^{x^\mp}_{-\infty} dx^\mp \frac{\rho_{(A,B)} (x^\mp, \vv x)}{\Delta_T - m^2} \bigg)},
\end{equation}
where $\Delta_T$ is the two-dimensional Laplace operator in the transverse plane and $m$ is an infrared regulator. The infrared regulator is used to model the phenomenon of  color confinement in the nucleus. It is common practice to choose the value of $m$ on the order of $\Lambda_\mathrm{QCD} \approx 200 \, \mathrm{MeV}$ in order to enforce color neutrality on transverse length scales of $\sim 1 \, \mathrm{fm}$. In the ultrarelativistic limit, where nuclei are assumed to be infinitesimally thin along $x^\mp$, the transverse gauge fields simplify to
\begin{align}
A^i_{(A,B)}(x^\mp, \vv x) &= \theta(x^\mp) \alpha^i_{(A,B)}(\vv x), \\
\alpha^i_{(A,B)}(\vv x) &= \frac{1}{ig} V( \vv x) \p^i V^\dg( \vv x),
\end{align}
where $\theta(x)$ denotes the Heaviside step function and the asymptotic Wilson lines are given by
\begin{equation} \label{eq:cgc_wilson_asy}
V^\dg_{(A,B)}(\vv x) = \lim_{x^\mp \rightarrow \infty} V^\dg_{(A,B)}(x^\mp, \vv x).
\end{equation}

To specify the charge densities $\rho_{(A,B)}$, a model has to be chosen. In this work, we focus on the McLerran-Venugopalan (MV) model \cite{MV1, MV2}. For a nucleus moving along $x^+$ (nucleus ``A''), the random charge density $\rho_A$ is assumed to be Gaussian and given by \cite{JalilianMarian:1996xn}
\begin{align}
&\ev{\rho^a_A(x^-, \vv x)} = 0, \\
&\ev{\rho^a_A(x^-, \vv x) \rho^b_A(y^-, \vv y)} = \nonumber \\
& \qquad g^2 \mu_A^2 \lambda_A(x^-) \delta^{ab} \delta(x^- - y^-)\delta^{(2)}(\vv x - \vv y), \label{eq:mv_twop}
\end{align}
where $g$ is the Yang-Mills coupling constant and $\mu_A$ is the MV model parameter given in units of energy. The function $\lambda_A(x^-)$ defines the shape of the nucleus along $x^-$ and is highly peaked around $x^- = 0$. Additionally, it is normalized, i.e.
\begin{equation}
\intop^{\infty}_{-\infty} dx^- \lambda_A(x^-) = 1.
\end{equation}
The expectation value $\ev{\dots}$ refers to the functional integration over all realizations of $\rho$, as specified by the probability functional $W_\mathrm{MV}[\rho]$, i.e.
\begin{equation}
\ev{\dots} = \int \mathcal{D} \rho \, W_\mathrm{MV}[\rho] \left(  \dots \right),
\end{equation}
where the functional $W_\mathrm{MV}[\rho]$ is given by
\begin{equation}
W_\mathrm{MV}[\rho] = Z^{-1} \exp{\left( - S_\mathrm{MV}[\rho] \right)},
\end{equation}
with
\begin{equation}
S_\mathrm{MV}[\rho] = - \frac{1}{2} \int dx^- \int d^2 \vv x \frac{\rho^a(x^-, \vv x) \rho^a(x^-, \vv x)}{g^2 \mu_A^2 \lambda_A(x^-)},
\end{equation}
and the normalization constant $Z$ is determined by
\begin{equation}
Z = \int \mathcal{D} \rho \,  \exp(- S_\mathrm{MV}[\rho]).
\end{equation}
We note that the MV model is both homogeneous and isotropic in the transverse plane as evident from eq.~\eqref{eq:mv_twop}. Consequently, nuclei described by the MV model have no finite radius. For a nucleus moving along $x^-$ (nucleus ``B''), the charge density correlator eq.~\eqref{eq:mv_twop} is defined analogously with $x^+$ in place of $x^-$.

The classical Glasma field, which is created in the collision of the two nuclei, is obtained from solving the classical Yang-Mills equations
\begin{equation}
D_\mu F^{\mu\nu} = J^\nu_A + J^\nu_B
\end{equation}
in the future light cone, which is spanned by the Milne coordinates: proper time $\tau = \sqrt{t^2 - z^2} \geq 0$ and space-time rapidity $\eta = \ln \left( (t + z) / (t - z) \right) / 2$. In general, there are no analytical solutions for the gauge field in the future light cone for arbitrary $\tau > 0$. It is however possible to deduce that gauge-invariant observables in the future light cone need to be invariant under boosts along the beam axis. This boost symmetry can be extended to gauge fields (without loss of generality) if rapidity dependent gauge transformations are disregarded. The fields in the future light cone therefore only depend on $\tau$ and $\vv x$, i.e.\ $A_\mu(x) = A_\mu(\tau, \vv x)$. 
Furthermore, using the assumption of boost invariance, it is possible to solve the Yang-Mills equations for the fields at the boundary of the future light cone at $\tau = 0$. The solution in temporal gauge $A^\tau = 0$, is given by \cite{Kovner:1995ja}
\begin{align}
A^i(0, \vv x) &= \alpha^i_A(\vv x) + \alpha^i_B(\vv x), \label{eq:ic_glasma1}\\
A^\eta(0, \vv x) &= \frac{ig}{2} \left[ \alpha^i_A(\vv x), \alpha^i_B(\vv x) \right], \label{eq:ic_glasma2}
\end{align}
and
\begin{align}
\p_\tau A^i(0, \vv x) &= 0, \\
\p_\tau A^\eta(0, \vv x) &= 0.
\end{align}
Taking the above expressions for $A^i$ and $A^\eta$ at $\tau = 0$ as initial data, we need to solve the temporal gauge Yang-Mills equations in Milne coordinates for $\tau > 0$:
\begin{align}
P^i &= \tau \p_\tau A_i, \label{eq:ym_eq1} \\
P^\eta &= \frac{1}{\tau} \p_\tau A_\eta, \label{eq:ym_eq2} \\
\p_\tau P^i &= \tau D_j F_{ji} - \frac{ig}{\tau} \left[ A_\eta, D_i A_\eta\right], \label{eq:ym_eq3} \\
\p_\tau P^\eta &= \frac{1}{\tau} D_i \left( D_i A_\eta \right). \label{eq:ym_eq4}
\end{align}
Here, the first two equations define the canonical momenta $P^i$ and $P^\eta$. The initial conditions in these variables are given by
\begin{align}
P^\eta(0, \vv x) &= - 2 A^\eta(0, \vv x), \\
A_i(0, \vv x) &= - A^i(0, \vv x),
\end{align}
with $A_\eta = P^i = 0$ at $\tau = 0$.

Given a solution to the Yang-Mills equations, we can evaluate the accumulated transverse momentum. Note that $x^+$ in eq.\ \eqref{eq:psquared_correlator_nonabelian_2} refers to $x^+ = (\tau + x) / \sqrt{2}$ as defined in eq.~\eqref{eq:lc1}. The accumulated momenta in $y$ and $z$ are given by
\begin{align} \label{eq:psquared_glasma1}
\ev{p^2_y(\tau)}_R &= \frac{g^2}{D_R} \intop_0^\tau d\tau' \intop_0^\tau d\tau'' \ev{ \Tr \left[ f^y(\tau') f^y(\tau'') \right]_R }, \\
\ev{p^2_z(\tau)}_R &= \frac{g^2}{D_R} \intop_0^\tau d\tau' \intop_0^\tau d\tau'' \ev{ \Tr \left[ f^z(\tau') f^z(\tau'') \right]_R }, \label{eq:psquared_glasma2}
\end{align}
where $f^y$ and $f^z$ are defined as 
\begin{align}
f^y(\tau) &= U(0, \tau) \left( \frac{1}{\tau} P^y(\tau) + F_{xy}(
\tau) \right) U(\tau, 0), \\
f^z(\tau) &= U(0, \tau) \left( P^\eta(\tau) + \frac{1}{\tau} D_x A_\eta(\tau) \right) U(\tau, 0).
\end{align}
Each of the terms in parentheses are understood to be evaluated along the lightlike trajectory $x^\mu(\tau) = (\tau, \tau, 0, 0)^\mu$. In temporal gauge, the Wilson lines reduce to
\begin{equation}
U(\tau, 0) = \mathcal{P} \exp{\bigg( - i g \intop_0^\tau d\tau' A_x (\tau') \bigg)},
\end{equation}
where $A_x(\tau)$ is evaluated at $x^\mu(\tau)$. In terms of chromo\-/electric and \-/magnetic field strengths, the functions $f^y$ and $f^z$ are given by
\begin{align}
f^y(\tau) &= U(0, \tau) \left( E_y(\tau) - B_z(\tau) \right) U(\tau, 0), \label{eq:fy_EB}\\
f^z(\tau) &= U(0, \tau) \left( E_z(\tau) + B_y(\tau) \right) U(\tau, 0). \label{eq:fz_EB}
\end{align}

The expectation values in eqs.~\eqref{eq:psquared_glasma1} and \eqref{eq:psquared_glasma2} refer to the medium average introduced in sec.~\ref{sec:formalism}. In the case of the Glasma, this expectation value corresponds to an average over the charge densities of both nuclei, i.e.
\begin{equation}
\ev{\dots} = \int \mathcal{D} \rho_A \mathcal{D} \rho_B \left( \dots \right) W_A[\rho_A] W_B[\rho_B].
\end{equation}

Due to homogeneity and isotropy of the MV model, the final results for $\ev{p_y^2(\tau)}$ and $\ev{p_z^2(\tau)}$ do not depend on the starting position of the particle. This would not be the case for more complicated initial conditions, which, for instance, account for the finite radii of nuclei. More generally, if $\mu^2_A$ in the charge density correlator eq.~\eqref{eq:mv_twop} is promoted to a function $\mu^2_A(\vv x)$ of the transverse coordinates $\vv x$, the results for $\ev{p^2_y(\tau)}$ and $\ev{p^2_z(\tau)}$ will depend on the initial transverse coordinate of the particle. 

It should also be noted that the initial conditions of the MV model naturally fulfill eqs.~\eqref{eq:langevin_force_1} and \eqref{eq:langevin_force_2}, since no single color component is preferred over any other. Therefore, no additional assumptions are necessary in order for eqs.~\eqref{eq:psquared_glasma1} and \eqref{eq:psquared_glasma2} to hold. 

There are two main problems with evaluating eqs.~\eqref{eq:psquared_glasma1} and \eqref{eq:psquared_glasma2}: first, it is difficult to obtain fully non-perturbative solutions to the Yang-Mills equations. Second, even if a solution $A_\mu$ is known,  eqs.~\eqref{eq:psquared_glasma1} and \eqref{eq:psquared_glasma2} are highly non-linear functionals of the gauge field due to the presence of the lightlike Wilson lines. In order to make progress, some approximations are necessary.

In the rest of this section we discuss two methods we can use to approximately compute $\ev{p^2_\perp}$. The first method is to assume that the fields of the nuclei are weak and therefore perturbative analytic solutions for the Glasma field can be obtained from a linearization of the Yang-Mills equations. The solution to the linearized equations can be plugged into eqs.~\eqref{eq:psquared_glasma1} and \eqref{eq:psquared_glasma2}, where contributions from the Wilson lines are negligible at leading order. The second method is to solve the non-perturbative Yang-Mills equations numerically on a lattice for the case of strong fields. This is done using real-time lattice gauge theory. Our expression for $\ev{p^2_\perp}$ can be evaluated numerically using the solution obtained from the lattice calculation. Although the weak field approximation is not as phenomenologically interesting as the fully non-perturbative strong field case, it provides a non-trivial check of the numerics for the lattice calculation. Additionally, the weak field approximation leads to a transparent interpretation of momentum broadening from the early stages of heavy ion collisions in terms of the properties of Glasma flux tubes.

\subsection{Weak field approximation}

In this section we first summarize the weak field approximation, which has been discussed more carefully in \cite{Kovner:1995ts}. Our main goal is to use the results of \cite{Kovner:1995ts} to find semi-analytical expressions for eqs.~\eqref{eq:psquared_glasma1} and \eqref{eq:psquared_glasma2}. 
A more detailed version of the calculation can be found in \cite{Schuh2019}.

\subsubsection{Weak field limit of the Glasma}

The weak field approximation is based on the assumption that the charge density of the nucleus is small and that the Coulomb field of the nucleus is therefore small as well. This allows us to treat the Glasma initial conditions perturbatively and the further time evolution of the Glasma, once it has been created, as effectively Abelian. We start by deriving the weak field approximation to the Glasma initial conditions.  Using the fact that $\rho$ is small, we expand the Wilson lines in eq.~\eqref{eq:cgc_wilson_asy} in powers of $g \rho$:
\begin{align}
V_{(A,B)}^\dg(\vv x) &= \one - i g \phi^{(1)}_{(A,B)}(\vv x) - \frac{g^2}{2} \phi^{(2)}_{(A,B)}(\vv x) \nonumber \\
& \qquad + \mathcal{O}\left( (g \rho_{(A,B)})^3 \right), 
\end{align}
where the potentials $\phi$ are related to the charge densities of the incoming nuclei via
\begin{align}
\phi^{(1)}_{(A,B)}(\vv x) &= \intop^{\infty}_{-\infty} dx^\mp\, \frac{\rho_{(A,B)} (x^\mp, \vv x)}{-\Delta_T + m^2}, \\
\phi^{(2)}_{(A,B)}(\vv x) &= \intop^{\infty}_{-\infty} dx^\mp \intop^{\infty}_{-\infty} dx'^\mp \, \mathcal{P}_{x^\mp} \bigg[ \left( \frac{\rho_{(A,B)} (x^\mp, \vv x)}{-\Delta_T + m^2} \right)  \nonumber \\
& \qquad \times \left( \frac{\rho_{(A,B)} (x'^\mp, \vv x)}{-\Delta_T + m^2} \right) \bigg].
\end{align}
In the second line $\mathcal{P}_{x^\mp}$ denotes path ordering along $x^\mp$. The transverse gauge fields of the nuclei up to quadratic order in $\rho$ are given by
\begin{align}
\alpha^i_{(A,B)} 
&= \frac{1}{ig} V_{(A,B)} \p^i V_{(A,B)}^\dg \nonumber \\
&= - \p^i \phi^{(1)}_{(A,B)} + \frac{ig}{2} \p^i \phi^{(2)}_{(A,B)} - ig \phi^{(1)}_{(A,B)} \p^i \phi^{(1)}_{(A,B)} \nonumber \\
& \qquad + \frac{1}{g} \mathcal{O}\left( (g \rho_{(A,B)})^3 \right).
\end{align}
Inserting this expression into eqs.~\eqref{eq:ic_glasma1} and \eqref{eq:ic_glasma2}, we can derive the weak field Glasma initial conditions:
\begin{align}
A^i(0, \vv x) &= -\p^i \left( \phi^{(1)}_{A} + \phi^{(1)}_{B} \right) + \frac{ig}{2} \p^i   \left( \phi^{(2)}_{A} + \phi^{(2)}_{B} \right) \nonumber \\
&\quad - i g \left( \phi^{(1)}_A \p^i \phi^{(1)}_A + \phi^{(1)}_B \p^i \phi^{(1)}_B \right) \nonumber \\
&\quad + \mathcal{O}(\rho_{(A,B)}^3), \\
A^\eta(0, \vv x) &= \frac{ig}{2} \left[ \p^i \phi^{(1)}_A, \p^i \phi^{(1)}_B \right] + \mathcal{O}(\rho_{(A,B)}^3).
\end{align}
In terms of field strengths, the Glasma at $\tau = 0$ corresponds to purely longitudinal chromo-electric and \-/magnetic fields \cite{Fries:2006pv, Lappi:2006fp}, $E_z = P^\eta$ and $B_z = - F_{xy}$. In the weak field limit, the initial values for $P^\eta$ and $B_z$ read
\begin{align}
B_z(0, \vv x) &= - i g \epsilon_{ij} \left[ \p_i \phi^{(1)}_A, \p_j \phi^{(1)}_B \right] + \mathcal{O}(\rho_{(A,B)}^3), \label{eq:Bz_weak} \\
E_z(0, \vv x) &= - i g \delta_{ij} \left[ \p_i \phi^{(1)}_A, \p_j \phi^{(1)}_B \right] + \mathcal{O}(\rho_{(A,B)}^3). \label{eq:Ez_weak}
\end{align}
Note that at leading order in the weak field expansion, the fields $E_z$ and $B_z$ do not contain quadratic $\phi^{(2)}$ terms, which implies that path ordering effects do not contribute at this order of the perturbative series. 

Provided that the initial field strengths are to be considered small, the further time evolution of the Glasma is approximately Abelian, see e.g.\ \cite{Kovner:1995ts} and \cite{Fujii:2008km}. A general solution to the linearized (Abelian) Yang-Mills equations \eqref{eq:ym_eq1} -- \eqref{eq:ym_eq4}, which is consistent with the Glasma initial conditions $P^\eta \neq 0$, $A_i \neq 0$, $P^i = A_\eta = 0$ at $\tau = 0$, is given in terms of the Green's functions:
\begin{align}
E_z(\tau, \vv x) &= \intop_k J_0(\left| \vv k \right| \tau) \tilde{E}_z(0, \vv k) e^{i \vv k \cdot \vv{x}}, \\
B_z(\tau, \vv x) &=  \intop_k  J_0(\left| \vv k \right| \tau) \tilde{B}_z(0, \vv k) e^{i \vv k \cdot \vv{x}},
\end{align}
where $\tilde{E}_z(0, \vv k)$ and $\tilde{B}_z(0, \vv k)$ are the Fourier components of the initial longitudinal field strengths and $J_\alpha(x)$ are Bessel functions of the first kind. We also use the shorthand $\int_k (\dots)$ for momentum integrals:
\begin{equation}
\intop_k (\dots) = \int \frac{d^2 \vv k}{(2 \pi)^2} (\dots). 
\end{equation}
The time evolution of the transverse field strengths $B_y = D_x A_\eta / \tau$ and $E_y = P^y / \tau$ in eqs.~\eqref{eq:fy_EB} and \eqref{eq:fz_EB} is given by
\begin{align}
E_y(\tau, \vv x) &= -\intop_k \frac{i k_x}{| \vv k |} J_1(|\vv k| \tau) \tilde{B}_z(0, \vv k) e^{i \vv k \cdot \vv{x}}, \\
B_y(\tau, \vv x) &= +\intop_k \frac{i k_x}{| \vv k |} J_1(|\vv k| \tau) \tilde{E}_z(0, \vv k) e^{i \vv k \cdot \vv{x}}.
\end{align}

\subsubsection{Momentum broadening in the dilute Glasma}

Having summarized the weak field approximation of the Glasma, we now apply it to compute the accumulated momenta of fast partons as derived in section \ref{sec:formalism}.
In the Abelian approximation, the Wilson lines can be dropped from eqs.~\eqref{eq:fy_EB} and \eqref{eq:fz_EB}. Inserting the above relations for the time dependent field strengths and evaluating them on the trajectory, we find
\begin{align}
f^y(\tau) &= - \intop_k \gamma(\tau, \vv k) \tilde{B}_z(0, \vv k), \\
f^z(\tau) &= + \intop_k \gamma(\tau, \vv k) \tilde{E}_z(0, \vv k),
\end{align}
where we define
\begin{equation}
\gamma(\tau, \vv k) = \left( \frac{i k_x}{|\vv k|} J_1(|\vv k| \tau) + J_0(|\vv k| \tau) \right) e^{i k_x \tau}.
\end{equation}
The accumulated momenta can be written as
\begin{align}
\ev{p^2_y(\tau)}_R &= \frac{g^2}{D_R} \intop^\tau_0 d\tau' \intop^\tau_0 d\tau'' \intop_{k'} \intop_{k''}   \gamma(\tau', \vv k') \gamma(\tau'', \vv k'') \nonumber \\
&\qquad \times  \ev{\Tr \left[ \tilde{B}_z(0, \vv k') \tilde{B}_z(0, \vv k'')\right]_R},
\end{align}
\begin{align}
\ev{p^2_z(\tau)}_R &= \frac{g^2}{D_R} \intop^\tau_0 d\tau' \intop^\tau_0 d\tau''\intop_{k'} \intop_{k''} 
\gamma(\tau', \vv k') \gamma(\tau'', \vv k'') \times \nonumber \\
& \qquad \ev{\Tr \left[ \tilde{E}_z(0, \vv k') \tilde{E}_z(0, \vv k'')\right]_R}.
\end{align}
Note that the only difference in $\ev{p^2_y}$ and $\ev{p^2_z}$ is due to the different initial electric and magnetic autocorrelation functions. The time evolution of the fields and the movement of the colored test particle through the Glasma is hidden in $\gamma(\tau, \vv k)$ and is the same for both $\ev{p^2_y}$ and $\ev{p^2_z}$. Since the correlators of $E_z$ and $B_z$ are explicitly evaluated at $\tau = 0$, the integrals over proper time only act on the function $\gamma(\tau, \vv k)$.
Using eqs.~\eqref{eq:Bz_weak} and \eqref{eq:Ez_weak} and performing the averages using the MV model, we find that the initial correlators in momentum space are given by
\begin{align}
& \ev{\Tr \left[ \tilde{B}_z(0, \vv k') \tilde{B}_z(0, \vv  k'')\right]_R} = \nonumber \\ 
&\qquad T_R D_F D_A \frac{Q_A^2 Q_B^2}{g^2}  (2 \pi)^2 \delta^{(2)}(\vv  k' + \vv k'') c_B(|\vv k'|, m), \\
& \ev{\Tr \left[ \tilde{E}_z(0, \vv k') \tilde{E}_z(0, \vv k'')\right]_R} =  \nonumber \\
& \qquad T_R D_F D_A \frac{Q_A^2 Q_B^2}{g^2} (2 \pi)^2 \delta^{(2)}(\vv k' + \vv k'') c_E(|\vv k'|, m),
\end{align}
with $Q_{(A,B)} = g^2 \mu_{(A,B)}$ and 
\begin{align}
c_B(|\vv k|, m) &= \intop_p \frac{(\vv p \times \vv k)^2}{(|\vv p|^2 + m^2)^2 (|\vv k - \vv p|^2 + m^2)^2}, \label{eq:cB_k} 
\end{align}
\begin{align}
c_E(|\vv k|, m) &= \intop_p \frac{(\vv p \cdot (\vv k - \vv p))^2 }{(|\vv p|^2 + m^2)^2 (|\vv k - \vv p|^2 + m^2)^2}. \label{eq:cE_k}
\end{align}
Here, we use $\vv p \times \vv k$ to denote the two-dimensional cross product in the transverse plane. 
Using the fact that $\gamma(\tau, -\vv k) = \gamma^*(\tau, \vv k)$, we find 
\begin{align} \label{eq:p_yz_weak_integrals}
&\ev{p^2_{(y,z)}(\tau)}_R = \nonumber \\ 
&\qquad Q_A^2 Q_B^2 \frac{T_R D_F D_A}{D_R} \intop_k  G(\tau, \vv k) c_{(B,E)}(|\vv k|, m),
\end{align}
where
\begin{equation} \label{eq:G_general}
G(\tau, \vv k) = \left| \intop^\tau_0 d\tau' \gamma(\tau', \vv k) \right|^2.
\end{equation}
Additionally, since $m$ is the only dimensionful scale that occurs in the integrals, we can re-parameterize both expressions and find
\begin{align} \label{eq:psquared_weak}
\ev{p^2_{(y,z)}(\tau)}_R &= \frac{Q_A^2 Q_B^2}{m^2} \frac{T_R D_F D_A}{D_R} d_{(B,E)} ( m \tau),
\end{align}
with the dimensionless function 
\begin{align} \label{eq:F_BE}
d_{(B,E)} ( m \tau) &= \intop_{k/m}  G(m \tau, \vv k / m) c_{(B,E)}(|\vv k|/ m, 1)  \nonumber \\
&= \intop_q G(\tilde{\tau}, \vv q) c_{(B,E)}(|\vv q|, 1),
\end{align}
with dimensionless proper time $\tilde{\tau} = m \tau$. 
Using the above expressions we can determine the accumulated transverse momenta as a function of $\tilde{\tau}$ by (numerically) solving the integrals in $d_{(E,B)} ( \tilde{\tau})$.
Numerical results are presented in section \ref{sec:results}.

\subsection{Lattice approximation for strong fields \label{sec:lattice}}

In this section we discuss our approach to compute the accumulated transverse momenta of fast partons in the Glasma using real-time lattice gauge theory methods. This allows us to numerically determine $\ev{p^2_{(y,z)}(\tau)}$ for non-perturbatively large Glasma fields, which are produced in relativistic heavy-ion collisions. 

\subsubsection{Real-time lattice gauge theory description of the Glasma}
The real-time lattice gauge theory approach is a finite difference method to numerically solve the Yang-Mills equations and is usually formulated the following way: we discretize the transverse plane on which the gauge field $A_\mu$ is defined as a regular, square lattice of size $N_T \times N_T$ with transverse lattice spacing $a_T$. The transverse length of the lattice is given by $L_T = N_T a_T$. The discrete time step is denoted as $\Delta \tau$ and the discrete proper times are given by $\tau_n = n \Delta \tau$. For convenience, we choose $\Delta \tau = a_T / n_\tau$, where $n_\tau \geq 2$ is an even integer. Requiring that $n_\tau$ is an even integer simplifies the lattice formulation of transverse momentum broadening in terms of test particles (see section \ref{sec:tmb_lattice}). The lower bound of $n_\tau = 2$ guarantees numerical stability of the finite difference scheme. 
As we use the MV model as our model for nuclei, we can employ periodic boundary conditions for the transverse lattice. 

Instead of the usual pair of fields and conjugate momenta, $(A_\mu, P^\mu)$, lattice gauge theory is formulated in terms of gauge links or link variables, which are the Wilson lines (usually in the fundamental representation) connecting the lattice sites of the regular lattice. The transverse gauge field components $A_i(\tau, \vv x)$ are replaced by the gauge links $U_{x, \hat{i}}(\tau_n)$, which connect the lattice site $x$ and the neighboring site $x + \hat{i}$ with $\hat{i}$ denoting the unit vector along the coordinate axis $x^i$. For small lattice spacing $a_T$ the gauge links are related to the usual gauge fields via
\begin{equation}
U_{x, \hat{i}}(\tau) \simeq \exp \left( i g a_T A_i(\tau, \vv x + \frac{a_T}{2} \hat{i} )\right).
\end{equation}
Gauge links starting at $x$ and going in the opposite direction of $\hat{i}$ are denoted by
\begin{equation}
U_{x, -\hat{i}}(\tau) = U^\dg_{x-\hat{i}, \hat{i}}(\tau).
\end{equation}
The conjugate momenta $P^i(\tau, \vv x)$ are replaced by $P^i_x(\tau)$ and are defined in the continuous time limit $\Delta \tau \rightarrow 0$ as
\begin{equation}
P^i_x(\tau) = - \frac{i \tau}{g a_T} \left( \p_\tau U_{x, \hat{i}}(\tau)\right) U^\dg_{x, \hat{i}}(\tau).
\end{equation}
As with the gauge links $U_{x,\hat{i}}$, the momenta $P^i_x$ are defined at the mid-point of the edge connecting $x$ and $x + \hat{i}$. The rapidity component of the gauge field $A_\eta(\tau, \vv x)$ is replaced by $A_{x, \eta}(\tau)$, which is defined at the lattice sites. The conjugate momentum $P^\eta_x(\tau)$ is defined via
\begin{equation}
P^\eta_x(\tau) = \frac{1}{\tau }\p_\tau A_{x, \eta}(\tau).
\end{equation}
Using this set of variables, the Yang-Mills equations \eqref{eq:ym_eq1} -- \eqref{eq:ym_eq4} are replaced by a leapfrog scheme for finite time steps $\Delta \tau > 0$ (see e.g.~\cite{Krasnitz:1998ns, Lappi:2003bi}): 
\begin{align}
A_{x,\eta}(\tau_{n+1}) &= A_{x,\eta}(\tau_{n}) + \Delta \tau \tau_{n+\frac{1}{2}} P^\eta_x(\tau_{n+\frac{1}{2}}), \\
U_{x,\hat{i}}(\tau_{n+1}) &= \exp \left( \frac{i g a_T \Delta \tau}{\tau_{n+\frac{1}{2}}} P^i_x(\tau_{n+\frac{1}{2}}) \right) U_{x,\hat{i}}(\tau_n), \\
P^\eta_x(\tau_{n+\frac{1}{2}}) &= P^\eta_x(\tau_{n-\frac{1}{2}}) + \frac{\Delta \tau}{\tau_n} \sum_i D^2_i A_{x,\eta}(\tau_n), \\
 P^i_x(\tau_{n+\frac{1}{2}}) & = P^i_x(\tau_{n-\frac{1}{2}}) \nonumber \\
&\mkern-64mu - \sum_j \frac{\Delta \tau \tau_n}{g a_T^3} \left[ U_{x,\hat{i}\hat{j}}(\tau_n) + U_{x,\hat{i}-\hat{j}}(\tau_n) \right]_{\mathrm{ah}} \nonumber \\
&\mkern-64mu - \frac{i g \Delta \tau}{\tau_n} \left[U_{x, \hat{i}}(\tau_n) A_{x+\hat{i},\eta}(\tau_n) U^\dg_{x, \hat{i}}(\tau_n), D^F_i A_{x,\eta}(\tau_n) \right],
\end{align}
where we define the anti-hermitian, traceless part of a matrix $X$ in the fundamental representation as
\begin{equation}
\left[ X \right]_\mathrm{ah} = \frac{1}{2i} \left( X - X^\dg\right) - \frac{1}{N_c} \Tr \left[ \frac{1}{2i} \left( X - X^\dg\right) \right]_F \, \one.
\end{equation}
The so-called plaquette variables $U_{x, \hat{i}\hat{j}}$ are $1 \times 1$ Wilson loops on the lattice and given by
\begin{equation}
U_{x, \hat{i}\hat{j}} = U_{x,\hat{i}} U_{x+\hat{i}, \hat{j}} U_{x + \hat{i} + \hat{j}, -\hat{i}} U_{x+\hat{j}, -\hat{j}}.
\end{equation}
Note that the conjugate momenta are defined at fractional time steps $\tau_{n+1/2}$, which allows the leapfrog scheme to be accurate up to quadratic order in $\Delta \tau$. The discrete analogues of the gauge covariant derivative $D_\mu$ are given by the forward and backward differences:
\begin{align}
D^F_i A_{x,\eta}(\tau_n) &= \frac{ U_{x,\hat{i}} A_{x+\hat{i},\eta} U^\dg_{x,\hat{i}} - A_{x,\eta}}{a_T}, \\
D^B_i A_{x,\eta}(\tau_n) &= \frac{ A_{x,\eta}- U_{x,-\hat{i}} A_{x-\hat{i},\eta} U^\dg_{x,-\hat{i}}}{a_T},
\end{align}
where all terms on the left are evaluated at proper time $\tau_n$.
The second derivative $D^2_i$ is defined as $D_i^2 A_{x,\eta}(\tau_n) = D^F_i D^B_i A_{x,\eta}(\tau_n)$, where no sum over $i$ is implied. 

The Glasma initial conditions have to be discretized on the lattice as well. The details of the discretized initial conditions have been worked out in detail in  \cite{Krasnitz:1998ns}. The starting point is to approximate the charge density correlator of the MV model (see eq.~\eqref{eq:mv_twop}) on a three-dimensional lattice. For simplicity we focus on a nucleus moving along $x^+$ without loss of generality.
We assume a rectangular shape $\lambda(x^-)$ along $x^-$ and split the nucleus into $N_s$ ``color sheets'' along the longitudinal direction, which allows us to account for path ordering along $x^-$ \cite{Fukushima:2007ki}. The discretized correlator then reads
\begin{equation} 
\ev{\rho^a_{x,m} \rho^b_{y,n}} = \frac{ g^2 \mu^2 }{N_s a^2_T} \delta_{mn} \delta^{ab} \delta_{xy},
\end{equation}
where the indices $m$ and $n$ refer to the different color sheets. The values $\rho^a_{x,m}$ of the lattice color charge density are uncorrelated, normally distributed random numbers.
After having generated a particular random configuration of color charges, we solve the discretized Poisson equation in the transverse plane for each individual color sheet:
\begin{align}
- \Delta_T \phi^{a}_{x,m} &= - \sum_{i=1,2} \frac{\phi^{a}_{x+\hat{i},m} + \phi^{a}_{x-\hat{i},m} - 2 \phi^{a}_{x,m}}{a_T^2} \nonumber \\
&= \rho^a_{x,m}.
\end{align}
The above equation can readily be solved in terms of Fourier coefficients. The solution reads 
\begin{equation}  \label{eq:poisson_fourier}
\tilde{\phi}^{a}_{k,n} = \frac{1}{\tilde{k}^2_T + m^2} \, \tilde{\rho}^a_{k,n},
\end{equation}
where the lattice momentum $\tilde{k}^2_T$ is given by
\begin{equation}
\tilde{k}^2_T = \left( \frac{2}{a_T} \right)^2 \sum_{i=1,2} \sin^2 \left( \frac{k_i a_T}{2} \right),
\end{equation}
and $m$ is the infrared regulator. Finally, the lightlike Wilson line can be approximated as a path ordered product of the individual sheets \cite{Fukushima:2007ki}:
\begin{equation}
V^\dg_{x} = \prod_{n=1}^{N_s} \exp \left( - i g \phi^{a}_{x,n} \tf T_F^a \right),
\end{equation}
where $\tf T^a_F$ are the generators in the fundamental representation of $\mathrm{SU}(N_c)$.
The transverse gauge links of the nucleus are given by
\begin{equation}
U^{(A,B)}_{x,\hat{i}} = V^{\vphantom{\dagger}}_{\vphantom{\hat{i}}(A,B),x} V^\dagger_{(A,B),x+\hat{i}}.
\end{equation}
The above outlined procedure is performed for each nucleus separately. 
The discrete analogues of eqs.~\eqref{eq:ic_glasma1} and \eqref{eq:ic_glasma2}, which provide the initial Glasma fields right after the collision at $\tau=0$, are given by \cite{Krasnitz:1998ns}
\begin{equation}  \label{eq:ic_lattice1}
\left[ \left( U^A_{x,i} + U^B_{x,i} \right)  \left( \one + U_{x,i} \left( \tau = 0 \right) \right)^\dg  \right]_{\mathrm{ah}}= 0,
\end{equation}
and
\begin{align}  \label{eq:ic_lattice2}
P^\eta_x( \tau = 0 ) &= \frac{1}{2g ( a^i )^2} \sum_{i=1,2} \big[ ( U_{x,i} - \one ) ( U^{B,\dg}_{x,i} - U^{A,\dg}_{x,i} ) \nonumber \\
& \quad + ( U^\dg_{x-i,i} - \one ) ( U^{B}_{x-i,i} - U^{A}_{x-i,i} ) \big]_\mathrm{ah},
\end{align}
with
\begin{align}
P^i_x ( \tau = 0 ) &= 0, \\
A_{x,\eta} ( \tau = 0 ) &= 0.
\end{align}
Note that eq.~\eqref{eq:ic_lattice1} has to be solved numerically, except for $N_c = 2$, where an analytic solution is known  \cite{Krasnitz:1998ns}.

\subsubsection{Lattice approximation for transverse momentum broadening of fast colored particles} \label{sec:tmb_lattice}

Having defined the initial conditions and the discrete field equations, the next step is to find discretizations of eqs.~\eqref{eq:fy_EB} and \eqref{eq:fz_EB}. In order to obtain a consistent procedure we take inspiration from the nearest grid point (NGP) scheme of the colored particle-in-cell (CPIC) method (see e.g.~\cite{Hu:1996sf, Moore:1997sn, Strickland:2007}). We imagine a particle with color charge $Q$ moving across the two-dimensional transverse lattice with trajectory $\vv x(\tau_n)$. As the particles' position changes with time, the color charge $Q$ of the particle contributes to the lattice charge density $\rho_x(\tau_n)$ at the lattice site $x$ which is nearest to $\vv x(\tau_n)$. Each time the nearest grid point of $\vv x(\tau_n)$ changes from $\vv x$ to e.g.\ $\vv x + \hat{i}$, the color charge $Q$ must be color rotated using the Wilson line connecting $\vv x$ and $\vv x + \hat{i}$, i.e.\ the gauge link $U_{x,\hat{i}}(\tau_n)$. In this example, the color rotation of the charge reads
\begin{equation}
Q(\tau_{n+1}) = U^\dg_{x,\hat{i}}(\tau_n) Q(\tau_n) U_{x,\hat{i}}(\tau_n).
\end{equation}
Color rotation of the charges is necessary in order to guarantee local gauge-covariant charge conservation. 

In this work, we only consider fixed, lightlike trajectories along the $x$-axis, i.e. $x^\mu(\tau_n) = (\tau_n, \tau_n, 0, 0)^\mu + x^\mu_0$, where $x_0$ denotes a starting lattice site at $\tau_0 = 0$ in the transverse plane. At time steps $t_n = n n_\tau \Delta \tau = n a_T$ the particle position ends up at one of the sites of the transverse lattice, denoted by $x_n$. The nearest grid point changes at times $\bar{t}_n = (n + 1/2) a_T$ at which point the charge undergoes color rotation with the gauge link connecting the appropriate edge on the lattice. The consecutive color rotations of the particles' charge amount to the lightlike Wilson lines in eqs.~\eqref{eq:fy_EB} and \eqref{eq:fz_EB}. We therefore use the approximation
\begin{align}
U(0, \bar{t}_n) &\approx U_{x_0, \hat{x}}(\bar{t}_0) U_{x_1, \hat{x}}(\bar{t}_1) \dots \nonumber \\
&\quad \times U_{x_{n-1}, \hat{x}}(\bar{t}_{n-1}) U_{x_n, \hat{x}}(\bar{t}_n).   
\end{align}
The field strengths $E_y$, $E_z$, $B_y$ and $B_z$ in eqs.~\eqref{eq:fy_EB} and \eqref{eq:fz_EB} acting on the particle are most easily computed at the lattice sites $x_n$ where the particle ends up at times $t_n$. The evaluation of the non-Abelian Lorentz force and the Wilson line are therefore not done at the same time. Rather, we alternate between computing field strengths at $t_n$ and the Wilson line at $\bar{t}_n$. 

We use the following approximations for the field strengths, which are accurate up to quadratic order in both time step $\Delta \tau$ and lattice spacing $a_T$:
\begin{align} 
E_y(t_n) &= \frac{1}{t_n} P^y(t_n) \nonumber \\
&\approx \frac{1}{4 t_n} \bigg( P^y_{x_n}(t_n + \frac{\Delta \tau}{2}) + P^y_{x_n}(t_n - \frac{\Delta \tau}{2})  \nonumber \\
&\qquad + U_{x_n, -\hat{y}}(t_n) \big[ P^y_{x_n-\hat{y}}(t_n + \frac{\Delta \tau}{2})\nonumber \\
& \qquad  + P^y_{x_n-\hat{y}}(t_n - \frac{\Delta \tau}{2}) \big] U^\dg_{x_n, -\hat{y}}(t_n) \bigg),
\end{align}
\begin{align} 
E_z(t_n) &= P^\eta(t_n) \nonumber \\
&\approx \frac{1}{2} \left( P^\eta_{x_n}(t_n + \frac{\Delta \tau}{2}) + P^\eta_{x_n}(t_n - \frac{\Delta \tau}{2}) \right),
\end{align}
\begin{align} 
B_y(t_n) &= - \frac{1}{t_n} D_x A_\eta(t_n) \nonumber \\
&\approx - \frac{1}{2 t_n a_T} \bigg( U_{x_n, \hat{x}}(t_n) A_{x_n + \hat{x}, \eta}(t_n) U^\dg_{x_n, \hat{x}}(t_n) \nonumber \\
& \qquad \qquad - U_{x_n, -\hat{x}}(t_n) A_{x - \hat{x}, \eta}(t_n) U^\dg_{x_n, -\hat{x}}(t_n) \bigg),
\end{align}
\begin{align} 
B_z(t_n) &= - F_{xy}(t_n) \nonumber \\
&\approx - \frac{1}{4 g a^2_T} \big[ U_{x_n,\hat{x}\hat{y}}(t_n) + U_{x_n,\hat{y}-\hat{x}}(t_n) \nonumber \\
& \quad \qquad + U_{x_n,-\hat{x}-\hat{y}}(t_n) + U_{x_n,-\hat{y}\hat{x}}(t_n)  \big]_\mathrm{ah}.
\end{align}

Finally, we put all the above expressions together to obtain the lattice approximation of eqs.~\eqref{eq:fy_EB} and \eqref{eq:fz_EB}:
\begin{align}
f^y(t_n) &= U(0, \bar{t}_n) \left( E_y(t_n) - B_z(t_n) \right) U(\bar{t}_n, 0), \label{eq:fy_EB_latt}\\
f^z(t_n) &= U(0, \bar{t}_n) \left( E_z(t_n) + B_y(t_n) \right) U(\bar{t}_n, 0), \label{eq:fz_EB_latt}
\end{align}
with the field strengths evaluated at $t_n$ and the Wilson line evaluated at $\bar{t}_n$. 

In order to compute the accumulated transverse momenta, we have to integrate over $\tau$, which we approximate using a simple sum
\begin{align}
\intop^\tau_0 d\tau' f^{(y, z)}(\tau') \approx a_T \sum^n_{i=0} f^{(y, z)}(t_n). 
\end{align}
The accumulated momenta on the lattice then read
\begin{align} \label{eq:psquared_latt}
\ev{p^2_{(y,z)}(t_n)}_R &
\approx\frac{g^2 a_T^2}{D_R} \ev{ \Tr \bigg[ 
\big(\sum^n_{i=0} f^{(y, z)}(t_n) \big)^2 \bigg]_R }.
\end{align}
The expectation value in the above expression is a simple average over the random color charge densities used in the initial conditions. The statistical quality of the average can be improved by additionally averaging over the starting positions $x_0$ of the particle. This is possible because the MV model used in this work is homogeneous in the transverse plane, which implies that the results must not depend on $x_0$ if one averages over enough configurations. The additional average over $x_0$ simply accelerates the convergence of the expectation value, thus reducing the computational cost of numerically determining eq.~\eqref{eq:psquared_latt}.

In practice, we actually evaluate eq.~\eqref{eq:psquared_latt} only for quarks, since the variables in our code are all in the fundamental representation. However, as shown in section \ref{sec:scaling}, a simple rescaling of our results yields the accumulated momenta for gluons. 

\section{Results and discussion \label{sec:results}}

In the first part of this section we study the dilute Glasma, where we are able to perform both semi-analytic calculations and pure lattice calculations. We find that fast partons in the dilute Glasma exhibit large anisotropic momentum broadening, which can be traced back to infrared behaviour of the initial chromo-electric and \-/magnetic fields in the Glasma. In the second part we study the dense Glasma by applying the lattice simulations to more realistic initial conditions. This allows us to estimate the effect of the Glasma on quark and gluon jets. We find that also in the case of the dense Glasma the effect of anisotropic momentum broadening persists.

The methods to compute the accumulated transverse momenta presented in the previous sections have been implemented as part of \caps{curraun} \footnote{The source code is publicly available at \href{https://gitlab.com/openpixi/curraun}{https://gitlab.com/openpixi/curraun}}, an open-source code for simulating the Glasma in 2+1D using GPUs written in Python and Numba. Jupyter notebooks are provided with the code which can be used to replicate the results presented in this section.

\subsection{The dilute Glasma}

\begin{figure}
    \includegraphics{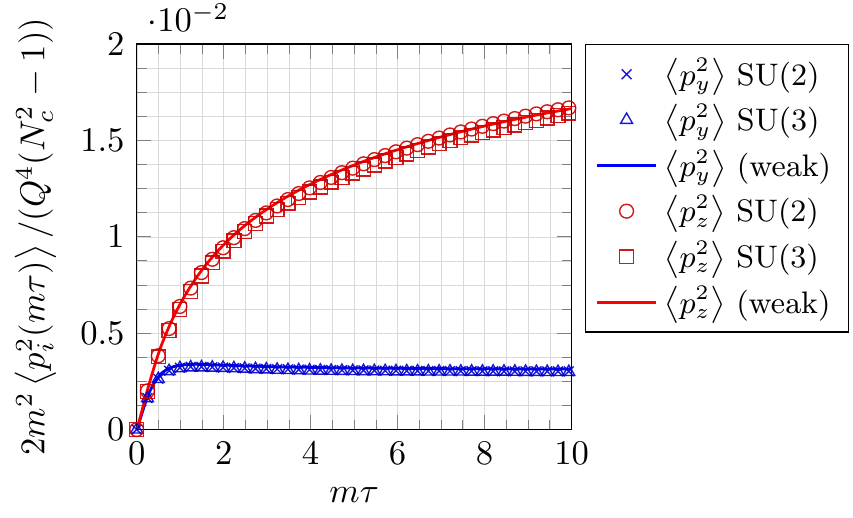}
    \caption{
    Accumulated transverse momenta in the dilute Glasma: weak field approximation vs.~lattice approximation.
    We performed simulations using both SU(2) (crosses and circles) and SU(3) (triangles and squares) to show that our lattice results scale with $D_A = N_c^2 - 1$. The results for the weak field approximation are shown as red and blue lines.
    It is evident that our lattice simulations applied to the dilute Glasma limit agree with the weak field approximation.
    While $\ev{p_z^2(m\tau)}$ increases monotonically with $m\tau$, $\ev{p^2_y(m\tau)}$ converges to a finite value after $m\tau > 1$. As a result, the two momentum components exhibit a large anisotropy which grows with $m \tau$.
    \label{fig:acc_mom_weak}}
\end{figure}

\begin{figure*}
\centering
\subfigure[Initial correlators in momentum space]{\includegraphics{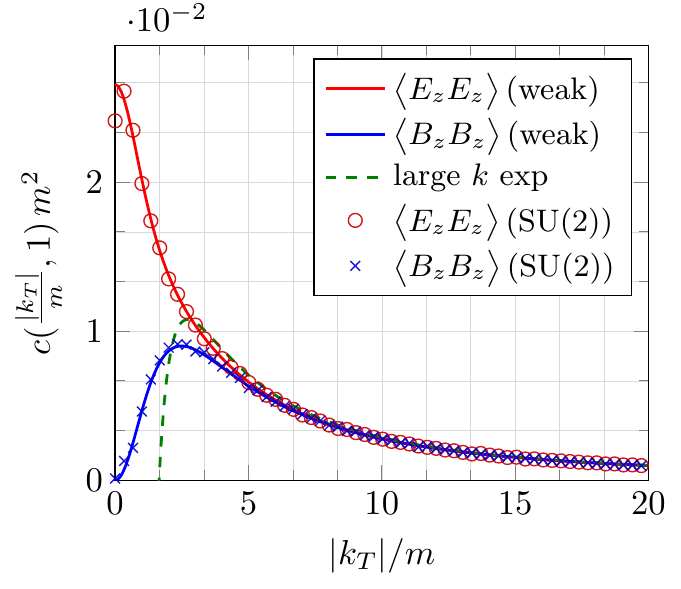}}
\quad
\subfigure[Initial correlators in coordinate space multiplied by the dimensionless distance $m r$ ]{\includegraphics{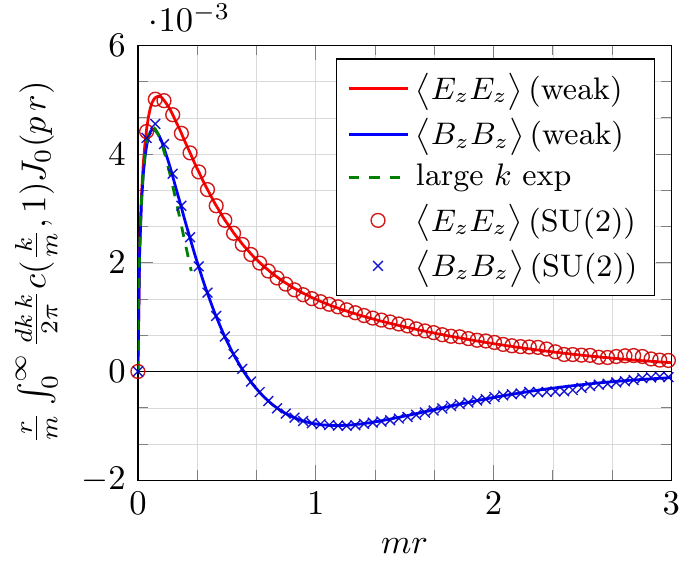}}

\caption{Correlators of initial longitudinal chromo-electric and \-/magnetic fields in the weak field approximation and the large $k$ expansion (see eq.~\eqref{eq:corr_large_k}). Solid lines refer to analytical results whereas circles and crosses are from a lattice simulation.
(a) A plot of eqs.~\eqref{eq:cB_k} and \eqref{eq:cE_k}. In momentum space, both correlators agree at high momenta, but their behaviour is different in the infrared region. At $k_T = 0$ the magnetic correlator vanishes while the electric correlator assumes a finite, non-zero value.
(b) A plot of eqs.~\eqref{eq:cB_k} and \eqref{eq:cE_k} after Fourier transformation in 2D polar coordinates with $r = \sqrt{x^2 + y^2}$. We see how the different infrared behaviour results in different correlation behaviour in coordinate space. While the electric field correlator is strictly positive everywhere, the magnetic field correlator exhibits a region of anti-correlation around  $r \approx m^{-1}$. At large distances $m r \gg 1$, both tend toward zero.
\label{fig:weak_correlators}}
\end{figure*}

We compute the accumulated transverse momentum in the weak field approximation eq.~\eqref{eq:psquared_weak} by numerically solving the integrals in eq.~\eqref{eq:F_BE}. We focus on collisions of identical nuclei and set $Q_A = Q_B = Q$. The dimensionless transverse momenta for quarks are then given by
\begin{align} \label{eq:acc_weak_dimless}
\frac{2 m^2}{Q^4 (N^2_c - 1)} \ev{p^2_{(y, z)} (m \tau)} &= d_{(B,E)}(m \tau) \nonumber \\
&= \intop_q G(\tilde{\tau}, \vv q) c_{(B,E)}(|\vv q|, 1).
\end{align}
We compute these integrals numerically using SciPy \cite{Virtanen:2019joe}.

The same quantities can be computed using our lattice simulation. We use a transverse lattice with $N_T = 1024$ cells per side and a time step of $\Delta \tau = a_T / n_\tau$ with $n_\tau = 16$. In order to approach the weak field limit we need to choose our infrared regulator $m$ to be much larger than $Q = g^2 \mu$. For the results presented here we used $m = 100 \, Q$. On the lattice we have to make sure that there is no interference from the cutoffs due to the lattice approximation, i.e. we make sure that $L_T ^{-1} \ll m \ll a_T^{-1}$, where $L_T^{-1} = (N_T a_T) ^{-1}$ is the infrared cutoff due to the finite size of the lattice and $a_T^{-1}$ is the ultraviolet cutoff due to the finite size of each lattice cell. In our simulations we chose $L_T = N_T a_T = 20 m^{-1}$ to satisfy this condition. For the discretization of the MV model we use $N_s = 50$ color sheets per nucleus. The results of eq.~\eqref{eq:psquared_latt} have been averaged over $20$ collision events. In order to show that $\ev{p^2_{(y,z)}} / (N^2_c - 1)$ is indeed independent of $N_c$, we performed our simulations with both SU(2) and SU(3). The results of both our lattice calculation and the weak field approximation are shown in fig.~\ref{fig:acc_mom_weak}, which show good agreement between our purely numerical lattice calculation and semi-analytical weak field approximation. We take this as evidence for a correct implementation of our lattice code.

The first important observation in fig.~\ref{fig:acc_mom_weak} is that although the accumulated momentum $\ev{p_z^2}$ along the beam axis keeps increasing with time, the $y$ component starts settling after $\tau \gtrsim m^{-1}$ to some constant value. The result is a large anisotropy $\ev{p^2_z} / \ev{p^2_y}$ which increases monotonically with time $\tau$. Consequently, the widening momentum cone around the particle trajectory becomes highly elliptical with much more pronounced broadening in $z$ or, equivalently, rapidity $\eta$. This anisotropy is rather interesting in the light of eq.~\eqref{eq:acc_weak_dimless}. The only difference between $\ev{p^2_y}$ and $\ev{p^2_z}$ is the type of correlator that is integrated over: either the initial magnetic correlator $c_B(| \vv k |/m, 1)$ or the electric correlator $c_E(| \vv k|/m, 1)$ (see eqs.~\eqref{eq:cB_k} and \eqref{eq:cE_k}). The term in eq.~\eqref{eq:acc_weak_dimless} which takes care of the time evolution of the system is hidden in $G(m \tau, k_T/m)$ and is the same for both momentum components. The interpretation is that in the dilute Glasma, a high energy parton receives momentum kicks in the $z$ direction only from the chromo-electric Glasma flux tubes, whereas the momentum kicks in the $y$ direction are due to chromo-magnetic flux tubes.
The physical reason for anisotropic momentum broadening is therefore not the longitudinal expansion of the Glasma as one might naively assume. The anisotropy emerges due to differences in the correlations of the initial chromo-electric and chromo-magnetic longitudinal fields. For example, it is a simple task to study a non-expanding Glasma by changing the $\tau$-dependent background metric $g_{\mu\nu} = \mathrm{diag}(1, -1, -1, -\tau^2)$ to the flat Minkowski metric $\eta_{\mu\nu} = \mathrm{diag}(1, -1, -1, -1)$. This only affects $G(m\tau, \vv k / m)$, but not the general form of eq.~\eqref{eq:acc_weak_dimless}. There exists a closed-form expression of $G(m\tau, \mathbf k / m)$ in the case of a flat background metric, given by
\begin{equation} \label{eq:G_flat}
G(m \tau, \mathbf k / m) = \frac{\sin(|\mathbf k| \tau)^2}{(|\mathbf k| / m)^2}
\end{equation}
Performing the calculation with eq.~\eqref{eq:G_flat} in place of eq.~\eqref{eq:G_general} leads to a similar anisotropic momentum broadening effect, although unaffected by the approximate free streaming expansion of the Glasma.
Isotropic momentum broadening is therefore possible only if the correlators of the chromo-electric and chromo-magnetic fields are the same. 
To be more precise: although the boost-invariant Glasma is anisotropic in the sense that it is a longitudinally expanding system and that at $\tau = 0$ the field strengths all point in the direction of the beam, the anisotropy observed in the momentum broadening is of a different physical origin and is related to the initial correlations in the Glasma.

In order to investigate the reason for this anisotropy more closely, we plot the autocorrelation functions of $E_z$ and $B_z$ at $\tau = 0$ in both momentum and coordinate space.  In the lattice approximation, we compute the gauge invariant field correlator, i.e.
\begin{align}
&C_E(|\vv x - \vv y|) = \left(  T_R D_F D_A \frac{Q_A^2 Q_B^2}{g^2} \right)^{-1} \nonumber \\
&\qquad \times \ev{\Tr \left[ E_z(\vv x) U_{\vv x  \rightarrow \vv y} E_z(\vv y) U_{\vv y  \rightarrow \vv x} \right]_R },
\end{align}
at $\tau = 0$, where $U_{\vv x  \rightarrow \vv y}(\tau)$ is the Wilson line connecting the lattice site $\vv x$ and $\vv y$. We include the Wilson line to guarantee gauge invariance, even though this only yields negligible corrections of higher order in the weak field limit. The momentum space correlator on the lattice is obtained by Fourier transforming the spatial correlator. The magnetic correlator is computed analogously.  The results are shown in fig.~\ref{fig:weak_correlators}.

First, focusing on the correlators in momentum space, we see that chromo-electric and \-/magnetic fields behave drastically differently in the infrared region $| \vv k| < m$. While the electric correlator admits a positive, non-zero value, the magnetic correlator tends to zero:
\begin{align}
\lim_{|\vv k| \rightarrow 0} \, c_E(|\vv k|, m) &= \frac{1}{12 \pi^2 m^2} > 0, \\
\lim_{|\vv k| \rightarrow 0} \, c_B(|\vv k|, m) &= 0.
\end{align}
On the other hand, for large momenta $|\vv k| \gg m$ both correlators agree:
\begin{align} \label{eq:corr_large_k}
c_E(|\vv k| \gg m, m) &\simeq c_B(| \vv k| \gg m, m) \nonumber \\
&\simeq \frac{1}{4 \pi |\vv k|^2} \left( \ln \left( \frac{| \vv k|^2}{m^2}\right) - 1\right).
\end{align}
The momentum broadening anisotropy of fast partons seems to be strongly affected by the infrared modes of the correlators. In the case of other observables that are less sensitive to the behaviour in the infrared, the mismatch in the two correlators for $|\vv k| < m$ can generally be neglected. For example, the energy density contributions of $E_z$ and $B_z$ at $\tau = 0$ are given by
\begin{align}
\varepsilon_{E}(\tau = 0) &= \frac{N_c (N_c^2 - 1) Q^4}{2 g^2} \intop^\Lambda_0 \frac{dk}{2 \pi} k \, c_{E}(k, m), \\
\varepsilon_{B}(\tau = 0) &= \frac{N_c (N_c^2 - 1) Q^4}{2 g^2} \intop^\Lambda_0 \frac{dk}{2 \pi} k \, c_{B}(k, m),
\end{align}
where $\Lambda \gg m$ is an ultraviolet cutoff. In the above expressions, the factor of $k$ from the integration measure suppresses the infrared modes of $c_E(|\vv k|, m)$ and $c_B(|\vv k|, m)$. Additionally, the integrand only falls off slowly with increasing $|\vv k|$ and therefore the main contributions to the (ultraviolet divergent) energy density come from high momenta $\vv k$, where $c_E(|\vv k|, m) \approx c_B(|\vv k|, m)$ and therefore $\varepsilon_{E}(\tau = 0) \simeq \varepsilon_{B}(\tau = 0)$ for $\Lambda \gg m$.

The fact that both electric and magnetic fields roughly yield the same initial energy density, combined with the Abelian time evolution of the Glasma, would suggest that there is, on average, no qualitative difference between a chromo-electric and a chromo-magnetic Glasma flux tube. However, this is only true at short distances. The different behaviour at larger length scales can be seen in the Fourier transformed correlators from the weak field approximation and our lattice simulation shown in fig.~\ref{fig:weak_correlators} (b).
For small distances $m r \ll 1$ both correlators and the large $k$ approximation of eq.~\eqref{eq:corr_large_k} agree. At intermediate distances $m r \approx 1$ however, the behaviour of the electric and magnetic correlators differs: the electric field correlator shows positive correlation for all $m r$, but the magnetic correlator has characteristic anti-correlated regions around $m r \approx 1$. Intuitively, this anti-correlation means that when looking at two points $\vv x_1$ and $\vv x_2$ in the transverse plane separated at a distance of $|\vv x_1 - \vv x_2| \approx m^{-1}$, it is very probable that the two magnetic field values $B_z(\vv x_1)$ and $B_z(\vv x_2)$ will have opposite signs. In contrast, the signs of $E_z(\vv x_1)$ and $E_z(\vv x_2)$ are more likely to be the same. The phenomenon of anti-correlated regions of the magnetic field at $\tau = 0$ has also been observed  in the dense Glasma \cite{Dumitru:2014nka}. 

In the dilute Glasma a physical interpretation of the zero mode of the correlators is given by the electric and magnetic flux in the transverse plane. We define the magnetic flux $\Phi_B(R)$ at $\tau = 0$ over a circular area of radius $R$ with origin $\vv x_0 = 0$ as
\begin{equation}
\Phi_B(R) = \intop_{C(R)} B_z(0, \vv x) \equiv  \intop_{C(R)} d^2 \vv x \, B_z(0, \vv x).
\end{equation}
Although the expectation value of $\Phi_B$ vanishes trivially, the expectation value of the trace of the magnetic flux with respect to the field at the origin, i.e. 
\begin{align}
&\ev{\Tr \left[ B_z(0, \vv x_0) \Phi_B(R) \right]_F} = \nonumber \\
&\intop_{C(R)} \intop_p \intop_q  e^{i (\vv x_0 \cdot \vv p + \vv x \cdot \vv q)} \ev{\Tr \left[ \tilde{B}_z(0, \vv p) \tilde{B}_z(0, \vv q)\right]_F} \nonumber \\
&= \frac{1}{2} N_c (N_c^2 - 1) \frac{Q^4}{g^2} \intop^\infty_0 dq \, R \,  J_1(q R) c_B(|\vv q|, m)
\end{align}
does not. In the limit of $R \rightarrow \infty$ we can use the fact that
\begin{equation}
\lim_{R \rightarrow \infty} R \, J_1(q R) = \delta(q)
\end{equation}
and find
\begin{align}  \label{eq:magnetic_flux}
&\lim_{R \rightarrow \infty }\ev{\Tr \left[ B_z(0, \vv x_0) \Phi_B(R) \right]_F} \nonumber \\
& \quad = \frac{N_c (N_c^2 - 1)}{2 g^2} c_B(|\vv 0|, m) = 0.
\end{align}
The vanishing of the zero mode implies that the total magnetic flux in the transverse plane vanishes. In contrast, the electric flux is non-zero:
\begin{align}  \label{eq:electric_flux}
&\lim_{R \rightarrow \infty }\ev{\Tr \left[ E_z(0, \vv x_0) \Phi_E(R) \right]_F} \nonumber \\
& \quad = \frac{N_c (N_c^2 - 1)}{2 g^2}c_E(|\vv 0|, m) = \frac{N_c (N^2_c - 1) Q^4}{12 \pi^2 m^2 g^2} > 0.
\end{align}

\begin{figure*}
\centering
\subfigure[Static Glasma with frozen flux tubes at $\tau = 0$]{\includegraphics{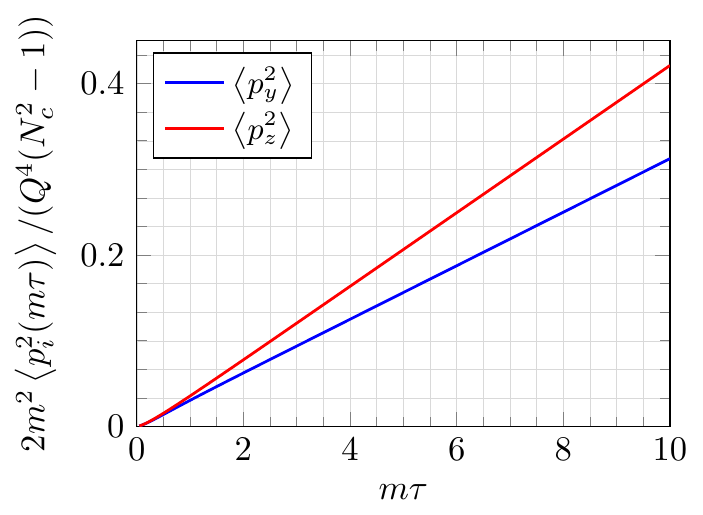}}
\quad
\subfigure[Evolving Glasma without longitudinal expansion]{\includegraphics{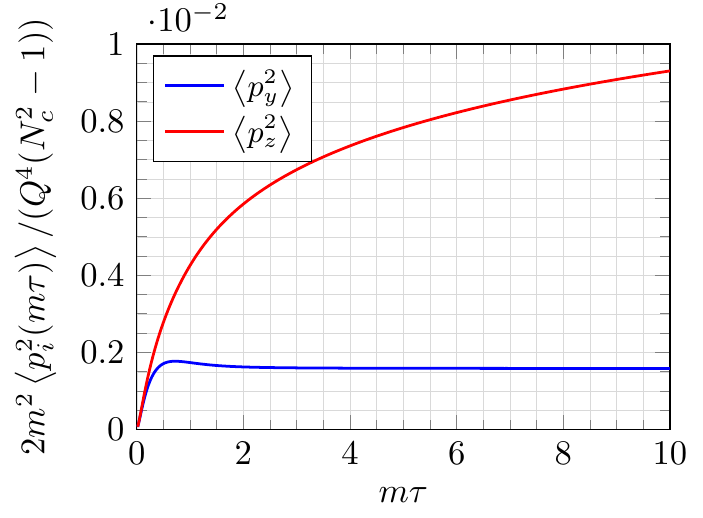}}

\caption{Accumulated transverse momenta for two scenarios: (a) the case of static Glasma flux tubes and (b) evolving Glasma flux tubes without longitudinal expansion. In (a) the Glasma flux tubes are assumed to be frozen in place at $\tau = 0$ and do not evolve according to the field equations. Due to differences in the correlations of the initial fields, see fig.~\ref{fig:weak_correlators}, more momentum is accumulated along $z$ compared to $y$.
In (b) the Glasma flux tubes evolve in a non-expanding background metric according to eq.~\eqref{eq:G_flat}. Comparing the two results it becomes apparent that the anisotropy is much more pronounced in the case of evolving flux tubes, which shows that the effect of partons co-moving with wave fronts is particularly important to explain the large anisotropy in the dilute Glasma.}
\label{fig:static_comp}
\end{figure*}

In the context of transverse momentum broadening of fast partons, the anti-correlation of the magnetic field (or the vanishing of the magnetic flux) has immediate consequences on the momentum broadening anisotropy. Consider a fast colored particle immediately after the collision that traverses the Glasma, which consists of electric and magnetic flux tubes. A fast particle that gets a particular momentum kick in the $y$ direction from a chromo-magnetic flux tube will, with high probability, get an anti-correlated momentum kick in the opposite direction after it has travelled a characteristic distance of $m^{-1}$, reducing the accumulated momentum $p_y$ of the particle. On the other hand, a particle that moves through an electric flux tube will not receive an anti-correlated kick in $z$ and therefore the accumulated momentum $p_z$ will be (on average) higher than $p_y$. This phenomenon can be seen clearly in fig.~\ref{fig:acc_mom_weak}, where it is shown that $\ev{p^2_y(\tau)}$ indeed flattens after a characteristic time of $m^{-1}$, corresponding to distances of $m^{-1}$ at the speed of light. In contrast, the other component $\ev{p^2_z(\tau)}$ does not exhibit such a decelerating effect as the electric field is always positively correlated. 

This picture is further complicated when taking into account that Glasma flux tubes evolve and expand in the $xy$-plane. As a fast parton is traversing the Glasma, it is affected, more or less randomly, by the wave fronts of flux tubes hitting the parton from all directions in the $xy$-plane. However, not all modes of the Glasma contribute equally to the accumulation of transverse momentum. In particular, modes that are co-moving with the particle (in this case, moving along the positive $x$ direction) are able to coherently accelerate the parton over a much longer time. As a result, the momentum broadening effect becomes even more sensitive to the long-ranged, infrared modes of the Glasma, which, as evident from fig.~\ref{fig:weak_correlators} (a), is very different for color-electric and \-/magnetic flux tubes (causing broadening in $p_z$ and $p_y$ respectively).
Within the weak field approximation, it can easily be demonstrated how co-moving modes in the Glasma affect the anisotropy of the momentum broadening effect by studying the case of a static Glasma, where all flux tubes produced at $\tau = 0$ are frozen in place and do not evolve. The results of this calculation are presented in fig.~\ref{fig:static_comp}, where it is shown that the anisotropy is much more pronounced in the case of the evolving Glasma. It is therefore a combination of the peculiar infrared behaviour of the initial flux tubes and the fact that the co-moving infrared modes are very efficient at accelerating the fast partons as they move through the Glasma that leads to the observed momentum broadening anisotropy. Interestingly, the case of static flux tubes shown in fig.~\ref{fig:static_comp} exhibits a linear increase of accumulated momenta $\ev{p^2_i(\tau)}$ with time $\tau$ for $m \tau \gtrsim 1$. It is possible to obtain the same result if one assumes an uncorrelated stochastic force $\mathcal{F}_i$ acting on the parton with
\begin{equation}
\ev{\mathcal{F}_i(\tau) \mathcal{F}_j(\tau')} \propto \delta_{ij} \delta(\tau - \tau').
\end{equation}
The scenario of static flux tubes is therefore highly similar to a two-dimensional random walk in transverse momentum space.

In summary, we find that anisotropic momentum broadening of fast partons moving through the dilute Glasma is closely linked to qualitative differences in the typical spatial shapes of chromo-electric and \-/magnetic Glasma flux tubes. Up until now we have only studied the dilute Glasma, which is of less phenomenological interest than the fully non-perturbative dense Glasma. The dilute Glasma, which neglects all non-linear interactions of the Yang-Mills field, should only be seen as a toy model for the dense gluonic matter created in relativistic heavy ion collisions. The dilute case is unphysical in the sense that the effective classical description in the CGC formalism is valid only for large fields and weak coupling. Moreover, the governing momentum scale in the weak field approximation is the infrared regulator $m$, which has to be introduced by hand in order to obtain finite results. The fact that all results depend strongly on this artificial energy scale is troubling. One might expect that the effect of anisotropic momentum broadening might be nothing but an artifact of the infrared regulation.
Fortunately, lattice gauge theory allows us to study the Glasma also in the case of strong fields, where the non-linear dynamics of the Yang-Mills fields seem to cure the infrared divergences that occur in the dilute limit and results should only weakly depend on the infrared regulator $m$. In particular, we will see that the governing momentum scale will be given by the saturation momentum $Q_s$ in the sense that our results for the accumulated momenta in the dense regime look qualitatively similar to fig.~\ref{fig:acc_mom_weak} if one replaces $m$ with $Q_s$.

\subsection{The dense Glasma: relativistic heavy ion collisions}

\begin{figure*}
\centering
\subfigure[Accumulated transverse momenta at early times]{\includegraphics{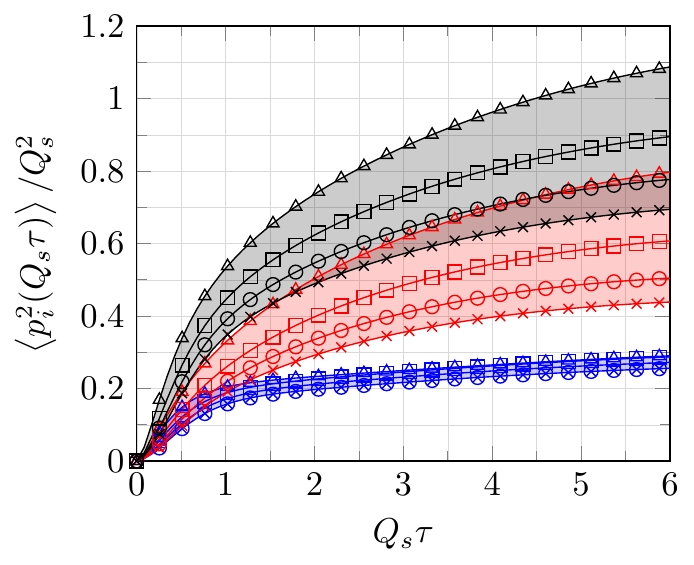}}
\quad
\subfigure[Late time behaviour of accumulated transverse momenta ]{\includegraphics{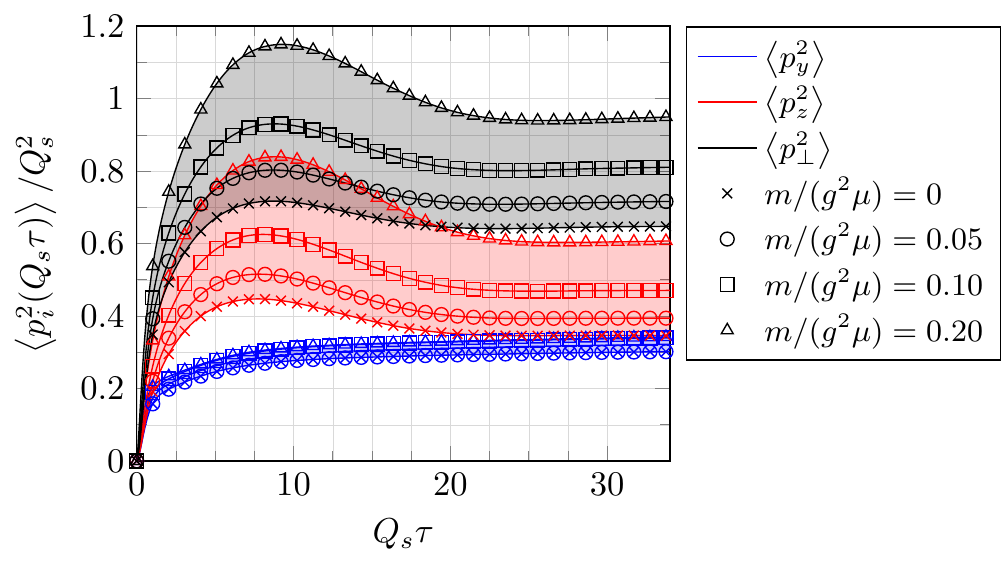}}

\caption{Accumulated transverse momentum for quarks as a function of dimensionless proper time $Q_s \tau$ from lattice simulations of the dense Glasma using SU(3) as the gauge group. The uncertainty in the results corresponds to different values of the infrared regulator $m$. In the case of $m=0$, the infrared modes are regulated by the system size $L$. The total squared transverse momentum (black lines and symbols) is given by $\ev{p^2_\perp} = \ev{p^2_y} + \ev{p^2_z}$. The results have been averaged over 50 random initial conditions. (a) At early times the behaviour of the two momentum components looks qualitatively similar to the dilute limit shown in fig.~\ref{fig:acc_mom_weak}: broadening in the $y$-component $\ev{p^2_y}$, which is initially determined by magnetic Glasma flux tubes, flattens out after $Q_s \tau \approx 1$. This can be traced back to a similar effect of anti-correlation at distances of $Q^{-1}_s$ in the longitudinal magnetic field at $\tau = 0$. Broadening along $z$, given by $\ev{p^2_z}$, is not affected by this. (b) At large time scales, at around $Q_s \tau \approx 10$, $\ev{p^2_y}$ starts to decrease again. This effect cannot be observed in the dilute Glasma. Additionally, $\ev{p^2_y}$ shows a strong dependence on $m / (g^2 \mu)$. For very small values of the infrared regulator, transverse momentum broadening becomes almost isotropic again at very late times. It should be noted that the classical effective approach of the Glasma is only valid up to $\tau \approx 10 \, Q^{-1}_s$.}
\label{fig:acc_mom_tau}
\end{figure*}

After having thoroughly studied momentum broadening in the dilute Glasma, we now focus on the dense Glasma, which is created in the collision of two relativistic heavy nuclei. As before, the color charge densities are generated according to the MV model. A dense Glasma  is produced when the ratio of the infrared regulator $m$ and the saturation momentum $Q_s$ (which is roughly $g^2 \mu$) becomes small: $m / (g^2 \mu) \ll 1$. 
On the other hand, the dilute Glasma corresponds to $m / (g^2 \mu) \gg 1$. 
To access the regime of non-perturbatively strong initial fields we perform lattice simulations as described in section \ref{sec:lattice}. In order to render our simulations more physical, we work at fixed saturation momentum $Q_s$. This done by choosing $g^2 \mu$ for some fixed $Q_s$ according the numerical results found in \cite{Lappi:2007ku}. A summary of the numerical parameters used in our simulations is shown in table \ref{tab:initial_parameters}.

\begingroup
\begin{table}
\caption{\label{tab:initial_parameters}
Parameters used in simulations of the dense Glasma. The numerical values for the ratios of $m$, $Q_s$ and $g^2 \mu$ were originally obtained by \cite{Lappi:2007ku}. The provided values correspond to $g^2 \mu L = 100$ with $N_s = 50$ color sheets per nucleus. For purposes of illustration, on the right side of the table we show example values for an intermediate saturation momentum of $Q_s = 1.5 \, \mathrm{GeV}$.
}
\begin{ruledtabular}
\begin{tabular}{cc|ccc}
$m / (g^2 \mu)$ & $Q_s / (g^2 \mu)$ & $Q_s$ & $g^2 \mu$ & $m$ \\
\colrule
0 & 1.08 & 1.5 GeV & 1.62 GeV & 0 GeV\\
0.05 & 0.98 & 1.5 GeV & 1.47 GeV & 0.08 GeV\\
0.10 & 0.85 & 1.5 GeV & 1.28 GeV & 0.15 GeV\\
0.20 & 0.68 & 1.5 GeV & 1.02 GeV & 0.30 GeV \\
\end{tabular}
\end{ruledtabular}
\end{table}
\endgroup

As in the case of the dilute Glasma, we work with square lattices of size $N_T = 1024$ and a time step of $\Delta \tau = a_T / 16$, where $a_T$ is the transverse lattice spacing. For $g^2\mu L = 100$, we achieve a lattice resolution of $g^2 \mu a_T \approx 0.1$. This is sufficient to resolve Glasma flux tubes which roughly have a diameter of $Q^{-1}_s \approx (g^2 \mu)^{-1}$ \cite{Lappi:2006fp}. For the infrared regulator $m$ we try out a few different values: $m / (g^2\mu) \in \{ 0, 0.05, 0.1, 0.2 \}$.
In the cases of $m \geq 0.05 g^2\mu$, the system size $L$ is large enough to resolve multiple color neutral domains with diameter $\sim m^{-1}$. In the case of $m / (g^2 \mu) = 0$ we simply eliminate the zero mode of the charge density (see eq.~\eqref{eq:poisson_fourier}) to implement color neutrality on the size of the system $L$ \cite{KRASNITZ2003268}. This corresponds to an effective infrared regulator of $m \propto L^{-1}$. 

Our main results are shown in fig.~\ref{fig:acc_mom_tau}, where we plot the accumulated transverse momenta for quarks in units of $Q_s$ as a function of dimensionless proper time $Q_s \tau$. For gluons the momenta should be multiplied with $C_A / C_F = 9 / 4$. There are a few observations we make: momentum broadening at early times $\tau \lesssim 5 / Q_s$ behaves remarkably similar to the dilute Glasma calculation, see fig.~\ref{fig:acc_mom_weak}. The broadening within the transverse plane $\ev{p^2_y}$ rises sharply until $\tau \approx Q_s^{-1}$, where it starts to flatten out. Broadening along the beam axis $\ev{p^2_z}$ is not affected by this and depends strongly on the infrared regulator $m$. We observe that the momentum broadening anisotropy starts to develop around $\tau \approx Q^{-1}_{s}$ and keeps growing until $\tau \approx 10 / Q_{s}$. The anisotropy shows strong dependence on the infrared regulator $m$ and shrinks for smaller values of $m$. At later times $\tau \gtrsim 10 / Q_s$ the large growth of $\ev{p^2_z}$ stops and is reversed. Surprisingly, the momenta along the beam axis are not broadened anymore but start to ``focus''. Consequently, the anisotropy shrinks and in the case of $m = 0$ vanishes almost entirely. One should keep in mind that the surprising behaviour of shrinking or almost vanishing momentum broadening anisotropy happens at rather late times. However, the classical approach is likely not valid after $\tau \approx 10 / Q_s$. It is therefore doubtful if this particular effect is of phenomenological interest. On the other hand, the early-time momentum broadening of fig.~\ref{fig:acc_mom_tau}~(a) is computed in the regime where the gluon occupation number of the Glasma is still large and a classical description can be applied. Therefore, we expect a similar behaviour of $\ev{p^2_y}$ and $\ev{p^2_z}$ also for more realistic CGC models of nuclei such as IP-Glasma, which, even though it accounts for non-homogeneity of the color charge density of nuclei due to varying nucleon positions, is at its core based on the MV model used in this work.  In fig.~\ref{fig:acc_mom_tau} we also plot the total accumulated momentum $\ev{p^2_\perp} = \ev{p^2_y} + \ev{p^2_z}$ as a function of $Q_s \tau$. We observe that after $\tau \approx 10 / Q_s$ a quark will have picked up roughly $\ev{p^2_\perp} \approx Q^2_s$. This value seems appropriate as the typical momentum of a gluon in the Glasma is roughly $Q_s$ \cite{Lappi:2011ju}.

\begin{figure}
    \includegraphics{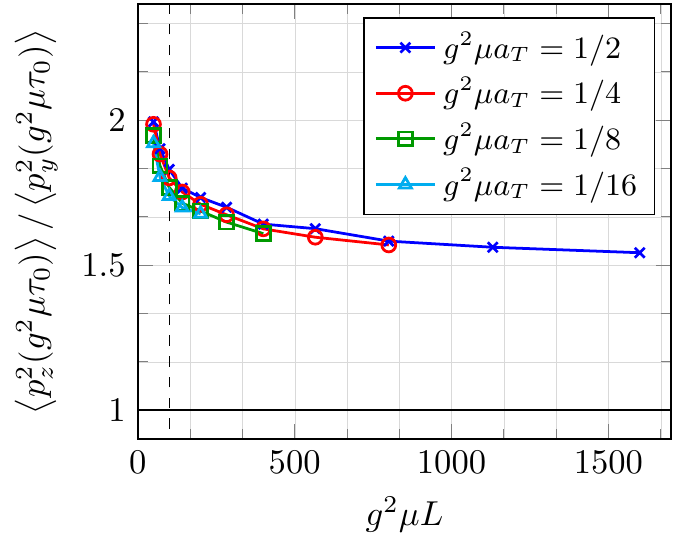}
    \caption{Momentum broadening anisotropy $\ev{p^2_z} / \ev{p^2_y}$ at $\tau_0 = 6 / (g^2 \mu)$ as a function of $g^2 \mu L$ for different transverse lattice spacings $g^2 \mu a_T$ from Glasma simulations using the gauge group SU(2). The dashed vertical line corresponds to the value $g^2 \mu L = 100$ used in the rest of this work. The anisotropy shows a dependence on $g^2\mu L$ for $g^2 \mu L < 500$, but this dependence becomes weaker for larger values of $g^2 \mu L$. Additionally, the anisotropy is not strongly affected by the lattice resolution $g^2 \mu a_T$. For each data point we performed an average over 200 random Glasma initial conditions.
    \label{fig:ani_large}}
\end{figure}

Looking at the infrared dependence of the momentum broadening anisotropy in fig.~\ref{fig:acc_mom_tau} (b) one might wonder if the large anisotropy around $Q_s \tau \approx 5 - 10$ still persists if the infrared regulator $m$ is reduced further. Recall that in the case of $m / (g^2 \mu) = 0$ the infrared divergence of the MV model is regulated by the system size $L^{-1}$. It is therefore possible to investigate the infrared dependence of $\ev{p^2_z} / \ev{p^2_y}$ for $m=0$ by studying the limit of large systems where $g^2 \mu L \rightarrow \infty$. These results are shown in fig.~\ref{fig:ani_large}. We simulate the Glasma using the gauge group SU(2) for fixed lattice spacing $g^2 \mu a_T$ and varying lattice size from $N_T = 100$ to $N_T = 3200$ (corresponding to the range of $g^2 \mu L = 100 g^2 \mu a_T$ to $g^2 \mu L = 3200 g^2 \mu a_T$). The results in fig.~\ref{fig:ani_large} have been obtained by averaging over 200 initial conditions for each value of $g^2\mu a_T$ and $g^2\mu L$. Although our results reveal a dependence of the anisotropy on $g^2 \mu L$ around the typical values used in the Glasma ($g^2 \mu L = 100$ as used for all other results in this work), the anisotropy never completely vanishes, i.e.~we find $\ev{p^2_z} / \ev{p^2_y} > 1.5$ also for large values of $g^2 \mu L$. It is important to note that we evaluate $\ev{p^2_z} / \ev{p^2_y}$ at fixed proper time $\tau_0 = 6 /(g^2 \mu)$, where the anisotropy is large according to fig.~\ref{fig:acc_mom_tau}. At asymptotically large $\tau$, the anisotropy almost vanishes. We conclude that while the accumulated momenta along the beam axis $\ev{p^2_z}$ are sensitive to the infrared regulator $m$ (or $L^{-1}$ in the case of $m = 0$), the momentum broadening anisotropy does not appear to be an artifact related to the way infrared modes are treated in the Glasma initial conditions. However, for $m \rightarrow 0$ there is only a short time window in which the anisotropy can become large. 

\begin{figure*}
\centering
\subfigure[Initial correlators in momentum space]{\includegraphics{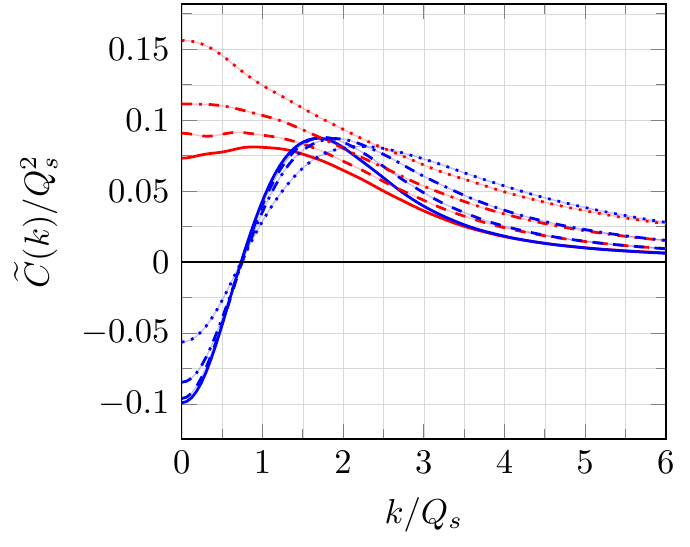}}
\quad
\subfigure[Initial correlators in coordinate space]{\includegraphics{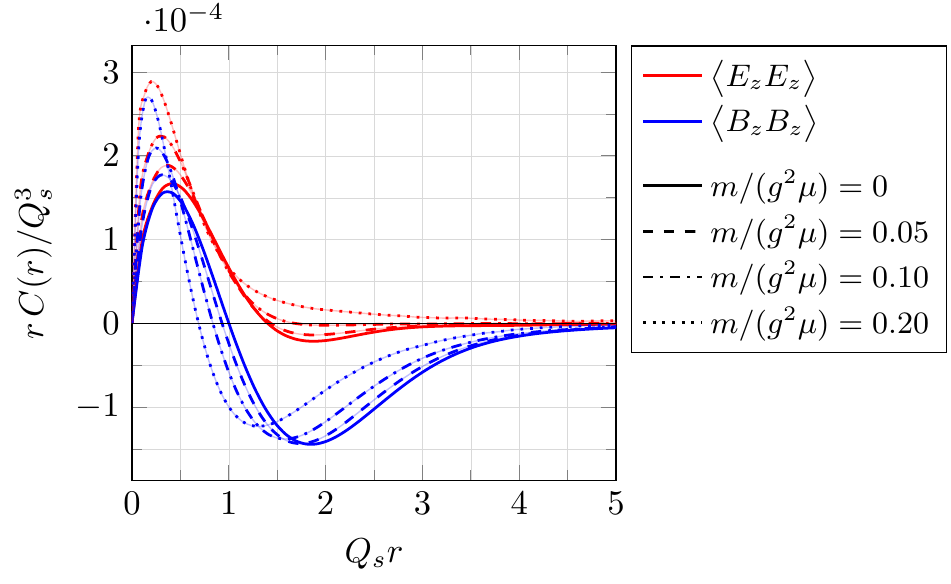}}

\caption{
Correlators of the initial field strengths in the dense Glasma case for various values of the infrared regulator $m$. (a) In momentum space, similar to the dilute case, the correlators of electric and magnetic fields agree at higher momenta, but show very different behaviour in the infrared: the zero mode of the magnetic correlator (which approximately corresponds to the expectation value of the magnetic flux in the dilute case, see eq.~\eqref{eq:magnetic_flux})
is negative. The zero mode of the electric correlator (which corresponds to electric flux in the dilute case) stays positive. As the ratio $m / (g^2 \mu)$ increases (moving more toward the dilute Glasma), the initial correlators start to resemble the weak field approximation results, see fig.~\ref{fig:weak_correlators} (a). 
(b) In coordinate space we observe a region of pronounced negative correlation in the magnetic field correlator, regardless of the infrared regulator $m$. For small values of $m$ the electric correlator remarkably also shows a small region of anti-correlation around $r \approx 2 \, Q^{-1}_s$. 
}
\label{fig:strong_correlators}
\end{figure*}

In order to further compare our non-perturbative results to the dilute Glasma and to better understand the time dependence of $\ev{p^2_y}$ and $\ev{p^2_z}$, we plot the gauge-invariant field strength autocorrelation functions at $\tau = 0$ from our lattice simulation
\begin{align}
C_E(|\vv x - \vv y|) &= \ev{\Tr \left[ E_z(\vv x) U_{\vv x \rightarrow \vv y} E_z(\vv y) U_{\vv y \rightarrow \vv x}  \right]}, \\
C_B(|\vv x - \vv y|) &= \ev{\Tr \left[ B_z(\vv x) U_{\vv x \rightarrow \vv y} B_z(\vv y) U_{\vv y \rightarrow \vv x}  \right]},
\end{align}
and their Hankel transformations
\begin{align}
\widetilde{C}_E(k) &= \intop^\infty_0 dr \, r \, C_E(r) J_0(k r), \\
\widetilde{C}_B(k) &= \intop^\infty_0 dr \, r \, C_B(r) J_0(k r),
\end{align}
in fig.~\ref{fig:strong_correlators}. As in the dilute Glasma case, we observe regions of negative correlation around $r \approx Q^{-1}_s$ in fig.~\ref{fig:strong_correlators}~(b). The same behaviour has been observed in \cite{Dumitru:2014nka}. We presume that this behaviour is linked to the flattening of $\ev{p^2_y}$ around $\tau \approx Q^{-1}_s$ in fig.~\ref{fig:acc_mom_tau}. It is interesting to note that in the dense Glasma the regions of anti-correlation are much more pronounced in comparison to the dilute limit. This is particularly apparent in the momentum space representation in fig.~\ref{fig:strong_correlators}~(a), where the zero mode of the magnetic field correlator is shown to be negative. As the ratio of the infrared regulator $m$ and $g^2 \mu$ increases, the amplitude of the zero mode approaches zero from below. In the limit of $m / (g^2 \mu) \rightarrow \infty$, which corresponds to the dilute limit, the zero mode vanishes as seen in fig.~\ref{fig:weak_correlators}~(b). Surprisingly, the electric field correlator also develops a region of anti-correlation for $m / (g^2 \mu) \lesssim 0.1$. This qualitative change from positive correlation for all distances $r$ to regions with positive and negative correlation in the color-electric field perhaps explains the strong dependence of $\ev{p^2_z}$ on the infrared regulator $m$, see fig.~\ref{fig:acc_mom_tau}.  It is clear that in the dense Glasma, as well as in the dilute case, the initial chromo-electric and \-/magnetic fields have very different properties. More intuitively, fig.~\ref{fig:strong_correlators} (b) suggests that the shapes of average electric and magnetic flux tubes differ. The average chromo-magnetic flux tube exhibits a strongly pronounced ``ring'' of anti-correlation around its center where the sign of the magnetic field differs from the field at the center. In contrast, for a typical chromo-electric flux tube this effect is much less pronounced.

Although we are not able to establish a simple relation between the accumulated momenta and the momentum space correlators as in the dilute case, see eq.~\eqref{eq:acc_weak_dimless}, we attribute the emergence of an anisotropy in transverse momentum broadening to the different correlation behaviour of the initial fields or rather the different shapes of chromo-electric and \-/magnetic flux tubes. Conversely, our calculation suggests that the observation of an anisotropy in the accumulated momenta of fast partons passing through the Glasma could be linked to differences in the properties of chromo-electric and \-/magnetic Glasma flux tubes.

\section{Conclusions and outlook}
In this paper we computed the transverse momentum broadening of highly relativistic partons as they traverse the boost-invariant Glasma created directly after the collision of two high energy nuclei. Using a test particle approximation, we derived simple, gauge invariant expressions for the accumulated momenta of colored particles in a non-Abelian background field using both the Wilson loop definition and classical colored particle dynamics.
This allowed us to study two interesting limits of the Glasma: the case of a dilute Glasma which can be described using the weak field approximation and the dense Glasma for which we have performed classical real-time lattice simulations.

In the case of the dilute Glasma we found that the accumulated transverse momenta can be directly computed from the correlation functions of the initial chromo-electric and \-/magnetic fields at $\tau = 0$. This leads to an intuitive picture of momentum broadening of fast particles in terms of the Glasma flux tubes: we observe that momentum broadening within the transverse plane is caused by the Lorentz force from chromo-magnetic flux tubes while momentum broadening along the beam axis is due to chromo-electric flux tubes. 
Using the MV model as an initial condition for the Glasma, we found a large momentum anisotropy with much stronger broadening of momenta along the beam axis compared to broadening within the transverse plane.
This anisotropy is caused by differences in the correlations of the initial fields: chromo-magnetic Glasma flux tubes exhibit a large region of anti-correlation which strongly suppresses the growth of momentum broadening within the transverse plane after a characteristic time scale. This stands in contrast to chromo-electric flux tubes, which show no signs of such anti-correlated regions and consequently fast partons can accumulate more momentum along the beam axis.
Remarkably, we also found that this momentum broadening anisotropy is not due to the longitudinal expansion of the system but stems purely from differences in the properties of the two types of Glasma flux tubes. Our real-time lattice simulation was also applied to the dilute case and found to be in very good agreement with the analytic results.

Finally, we studied the case of a dense Glasma using real-time lattice gauge theory which is phenomenologically more relevant for high energy nuclear collisions. We find that fast quarks accumulate up to $\ev{p^2_\perp} \approx Q^2_s$ of total transverse momentum within proper times of $\tau_0 \approx 6 -  10 \, Q^{-1}_s$. Here, $\tau_0$ roughly corresponds to the switching time from the Glasma description to a hydrodynamical stage \cite{Gale:2012rq, Romatschke:2017ejr}. Translating this into appropriate units we can use saturation momenta in the range of $Q_s = 1 \, \mathrm{GeV}$ to $Q_s = 2 \, \mathrm{GeV}$. Highly relativistic quarks acquire total accumulated transverse momenta of $\ev{p^2_\perp} \approx 1 \, \mathrm{GeV}^2$ to $\ev{p^2_\perp} \approx 4 \, \mathrm{GeV}^2$ after traversing the Glasma for proper times of $\tau_0 \approx 1.2 - 2 \, \mathrm{fm} /c$ and $\tau_0 \approx 0.6 - 1 \, \mathrm{fm} /c$ respectively. For highly energetic gluons moving through the Glasma, the accumulated (squared) momenta are to be multiplied by a factor of $C_A / C_F = 9 / 4$. We highlight that our main results shown in fig.~\ref{fig:acc_mom_tau} allow us to estimate $\ev{p^2_\perp}$ for a large range of values for $Q_s$, proper times $\tau$ and different values of the infrared regulator $m$.

Similar to the dilute Glasma, we find that momentum broadening in the dense Glasma is anisotropic with larger broadening along the beam axis. Here, we see that in particular the accumulated momentum along the beam axis $\ev{p^2_z}$ depends strongly on the infrared regulator $m$ used in the initial conditions. In order to rule out that this anisotropy is simply an artifact of infrared regulation, we studied the limit of very large systems with vanishing infrared regulator and found that the anisotropy still persists.

We also studied the field strength correlation functions in the dense Glasma and found similar behaviour as in the dilute case. Chromo-magnetic flux tubes at $\tau = 0$ exhibit pronounced regions of strong anti-correlation around $r \approx Q^{-1}_s$. Chromo-electric flux tubes on the other hand show much smaller anti-correlation around a larger length scale of $r \approx 2\, Q^{-1}_s$. Presumably, these qualitative differences in the two types of Glasma flux tubes are causing the momentum broadening anisotropy. By reversing this argument we conjecture that highly energetic partons are able to probe differences in the properties of chromo-electric and \-/magnetic fields of the Glasma. Anisotropic momentum broadening of fast partons with stronger growth of momentum along the beam axis has also been found in the quark-gluon plasma with momentum anisotropy (see \cite{Romatschke:2006bb} and \cite{2012JHEP...07..031G, 2012JHEP...08..041C, Rebhan:2012bw}). Our findings suggest that anisotropic broadening already occurs before the quark-gluon plasma phase. More work is necessary to see if these similar effects from two different stages of heavy ion collisions can be disentangled. 
It should be noted that the simple relation between the momentum broadening anisotropy and the correlators of initial fields has been derived only in the weak field approximation. Nevertheless, our simulations in the non-perturbative regime seem to suggest a similar relationship.

The methods presented in this paper can be extended in a number of interesting ways. There are two main directions that would improve our results: by using more realistic initial conditions for the Glasma and by using more accurate approximations for the interaction between the Glasma field and the partons moving through it. 

The MV model used in this work is lacking a number of features that more sophisticated nuclear models such as IP-Glasma \cite{Schenke:2012wb, Schenke:2012fw} could provide to obtain more accurate estimations of jet momentum broadening from the pre-equilibrium stage. For example, the color charge density in the MV model is homogeneous on average while IP-Glasma accounts for variations due to the spatial distribution of nucleons within a nucleus. Additionally, the MV model as it was used in this work can only be applied to central collisions (i.e.~zero impact parameter) and jets created in the center of the Glasma as there is no notion of a finite nuclear radius. Clearly, a relativistic parton created near the border of the Glasma, where the field strengths are much weaker compared to the center of the Glasma, would acquire less momentum from its interaction with the Glasma. Similarly, in off-central collisions the fields in the Glasma are much weaker leading to less momentum broadening overall. The MV model allows us to average over the starting position of the jet in the Glasma, but realistic initial conditions would require a more careful treatment of where and how particles are created in the collisions of high energy nuclei. It is also likely that near the border of the Glasma the transverse flow of the medium modifies the momentum broadening anisotropy (see \cite{Armesto:2004pt} for details of such an effect in a hydrodynamical setting). 

The calculation presented here is performed within the framework of a purely classical Yang-Mills field theory and does not account for corrections to the CGC initial conditions due to quantum fluctuations. To remedy this it would be necessary to use initial conditions evolved using the JIMWLK equations \cite{Iancu:2000hn, Ferreiro:2001qy}. The JIMWLK evolution allows an effectively classical description of color sources including quantum fluctuations, which can then be used as Glasma initial conditions in place of the original MV model initial conditions.
Earlier studies \cite{Lappi:2011ju} showed that the gluon spectrum in the Glasma becomes harder (i.e.~is shifted toward the ultraviolet) due to the JIMWLK evolution, leading to a larger average momentum per gluon in the Glasma.
This could increase the amount of momentum accumulated by a jet as it moves through and interacts with the Glasma.
Moreover, the JIMWLK evolution also allows for a rapidity dependent description of the Glasma \cite{Schenke:2016ksl}, which would extend the purely boost invariant approach of this work.

Another way to remove the assumption of boost invariance is to account for the finite longitudinal width of the colliding nuclei along the beam axis. 
Numerical approaches to simulate this genuinely 3+1 dimensional scenario have been developed by the authors previously \cite{Gelfand:2016yho, Ipp:2017lho, Ipp:2018hai}. The methods for computing jet momentum broadening developed in this paper could be applied to the 3+1D case in a straightforward manner.

All of the above noted extensions are focused on changing the non-Abelian background field, but it is also important to improve the approximations used for the interaction of the particle and the background field. In this paper we approximated highly energetic partons as colored test particles moving at the speed of light, i.e.~at infinitely high initial momentum. As a consequence of this approximation, the trajectory of the test particle does not change even though it acquires momentum along its path. 
Our results for momentum broadening therefore do not depend on initial momentum and should be interpreted as a limiting case of a more general calculation for finite initial momenta. The most obvious disadvantage of this approximation is that it does not account for energy loss of particles traversing the Glasma.
A partial solution to this is to include perturbative, non-eikonal corrections by allowing for small deviations from the straight line trajectory. More generally, a simulation of dynamic test particles using the colored particle-in-cell (CPIC) method \cite{Hu:1996sf, Moore:1997sn, Strickland:2007} could be used to numerically determine the full test particle trajectories. Going beyond the test particle approximation, it would also be interesting to include back reaction from the Glasma: as a colored particle moves through the Glasma, the Coulomb field of the particle interacts with the Yang-Mills field of the Glasma. As a consequence, the Glasma around the colored particle becomes polarized. This polarization leads to an induced chromo-electric field which 
would cause energy loss of the particle (see e.g.~\cite{Thoma:1990fm, Romatschke:2006bb}). The effect of parton energy loss from polarization effects in the Glasma could be computed using linear response \cite{Kurkela:2016mhu, Boguslavski:2018beu} or also using a fully dynamical CPIC setup.

\section{Acknowledgement}
We thank Carlos A.~Salgado for suggesting to look into this topic and Kirill Boguslavski, Tuomas Lappi and Jarkko Peuron for helpful discussions. We also thank Anton Rebhan and Alexander Soloviev for comments regarding the manuscript. 
This work has been supported by the Austrian Science Fund FWF No.~P32446-N27 and No.~P28352. 
The Titan\,V GPU used for this research was donated by the NVIDIA Corporation.

\appendix
\section{Taylor expansion of the Wilson loop} \label{sec:app_wilson}

In this appendix we give a detailed derivation of the accumulated squared momentum $\ev{p^2_i(t)}_R$ in eq.~\eqref{eq:psquared_correlator_nonabelian} using the Wilson loop definition in a non-Abelian background field. Analogous to the derivation in the Abelian case (see section \ref{sec:tmb_abelian}), we consider a rectangular Wilson loop with the transverse extent aligned with the $y$ axis and the lightlike extent aligned with $x^+$ (see fig.~\ref{fig:wilson_loop}). The Wilson loop is given by
\begin{equation} \label{eq:nonabelian_loop}
W_{y+} = W_{+, 2} W_{y, 2} W_{+, 1} W_{y, 1},
\end{equation}
where the lightlike Wilson lines are
\begin{align}
W_{+, 1} &= \mathcal{P} \exp{\bigg( - i g \intop_0^{L^+} dx^+ A_+(x^+, L) \bigg)}, \label{eq:na_ll_w_1} \\
W_{+, 2} &= \mathcal{P} \exp{\bigg( - i g \intop_{L^+}^0 dx^+ A_+(x^+, 0) \bigg)}, \label{eq:na_ll_w_2}
\end{align}
and the transverse Wilson lines are
\begin{align}
W_{y, 1} &= \mathcal{P} \exp{\bigg( - i g \intop_0^{L} dy A_y(0, y) \bigg)}, \label{eq:na_trans_w_1}\\
W_{y, 2} &= \mathcal{P} \exp{\bigg( - i g \intop_L^0 dy A_y(L^+, y) \bigg)}. \label{eq:na_trans_w_2}
\end{align}
Here, $\mathcal{P}$ denotes the path ordering symbol and $g$ is the Yang-Mills coupling constant. The path ordering symbol orders terms according to their occurrence along the path. Quantities that come ``later'' in the path are to the left of quantities that come ``before''. For readability we suppress the dependence of $A_\mu$ on $x^-$ and $z$. The definition of the Wilson loop $W_{z+}$ with the transverse extent in the $z$ direction is completely analogous. 

In order to perform the Taylor expansion in $L$  we have to expand three of the four Wilson lines in eq.~\eqref{eq:nonabelian_loop} up to linear order in $L$ to obtain $W^{(1)}_{y+}$. The Taylor series expansions are given by
\begin{align}
W_{y,1} &= \one + L W^{(1)}_{y,1} + \mathcal{O}(L^2), \\
W_{y,2} &= \one + L W^{(1)}_{y,2} + \mathcal{O}(L^2), \\
W_{+,1} &= W^{(0)}_{+,1} + L W^{(1)}_{+,1} + \mathcal{O}(L^2).
\end{align}
Note that the zeroth order term $W^{(0)}_{+,1}$ is given by the inverse of $W_{+,2}$, i.e. 
\begin{equation}
    W^{(0)}_{+,1} = W^\dg_{+,2}.
\end{equation}
The linear coefficient of the Taylor series is therefore
\begin{align}
W^{(1)}_{y+} =& W_{+,2} W^{(1)}_{y,2} W^\dg_{+,2} + W_{+,2} W^{(1)}_{+,1} + W^{(1)}_{y,1}.
\end{align}
The linear coefficients of the transverse Wilson lines are
\begin{align}
W^{(1)}_{y,1} &= - i g A_y(0, 0), \\
W^{(1)}_{y,2} &= + i g A_y(L^+, 0).
\end{align}
The expansion of the lightlike Wilson line $W_{+,1}$ is slightly more involved. The linear coefficient reads
\begin{align}
W^{(1)}_{+,1} = &- i g \intop_0^{L^+} dx'^+ W_+(L^+, x'^+)  \nonumber \\ 
&\qquad \times \p_y A_+(x'^+, 0) W_+(x'^+, 0),
\end{align}
where $W_+(b, a)$ is a lightlike Wilson line from $x^+=a$ to $x^+=b$ given by
\begin{equation}
W_+(b, a) = \mathcal{P} \exp{\bigg( - i g \intop_a^{b} dx^+ A_+(x^+, 0) \bigg)}.
\end{equation}
Putting everything together, the linear coefficient of $W^{(1)}_{y+}$ reads
\begin{align} \label{eq:W_linear_1}
W^{(1)}_{y+} &= i g \left( W_{+, 2} A_y(L^+, 0) W^\dg_{+,2} - A_y(0,0) \right) \nonumber \\
&\quad - i g  \intop_0^{L^+} dx^+ W_+(0, L^+) W_+(L^+, x^+) \nonumber \\ 
&\qquad \times \p_y A_+(x^+, 0) W_+(x^+, 0).
\end{align}
Note that
\begin{equation}
W_{+,2} = W_+(L^+, 0)
\end{equation}
The first term in eq.~\eqref{eq:W_linear_1} can be written as an integral over $x^+$:
\begin{align}
& \qquad W_{+, 2} A_y(L^+, 0) W^\dg_{+,2} - A_y(0,0) \nonumber \\
 &= \intop_0^{L^+} dx^+ \p_+ \left(  W_+(0, x^+) A_y(x^+, 0) W_+( x^+, 0) \right) \nonumber \\
&= \intop_0^{L^+} dx^+ W_+(0, x^+) \bigg( \p_+ A_y(x^+, 0) \nonumber \\
& \qquad \quad + ig \left[ A_+(x^+, 0), A_y(x^+, 0) \right] \bigg) W_+(x^+, 0),
\end{align}
where we have made use of
\begin{equation}
\p_+ W_+(x^+, 0) = - i g A_+(x^+, 0) W_+(x^+, 0).
\end{equation}
Writing everything in eq.~\eqref{eq:W_linear_1} under one integral yields
\begin{align} \label{eq:W_linear_2}
W^{(1)}_{y+} &= i g \intop_0^{L^+} dx^+ W_+(0, x^+) F_{+y}(x^+, 0) W_+(x^+, 0). 
\end{align}
Next, we define the parallel transported field strength as
\begin{align} \label{eq:parallel_transported_F_app}
\widetilde{F}_{y+}(x^+, 0) &= W_+(0, x^+) F_{y+}(x^+, 0) W_+(x^+, 0)
\end{align}
and rewrite the expression for the linear coefficient:
\begin{align} \label{eq:W_linear_3}
W^{(1)}_{y+} &= i g \intop_0^{L^+} dx^+ \, \widetilde{F}_{y+}(x^+, 0). 
\end{align}
The accumulated squared momentum (see eq.~\eqref{eq:psquared_linear_nonableian})
\begin{align}
\ev{p^2_i(t)}_R &= - \frac{1}{D_R} \ev{\Tr \left[ \left( W^{(1)}_{i+} \right)^2 \right]_R}
\end{align}
then becomes
\begin{align} \label{eq:psquared_correlator_nonabelian_y_app}
&\ev{p^2_y(t)}_R = \nonumber \\
&\quad \frac{2 g^2}{D_R} \intop_0^t dt' \intop_0^t dt'' \ev{\Tr \bigg[  \widetilde{F}_{y+}(x(t')) \widetilde{F}_{y+}(x(t''))\bigg]_R },
\end{align}
where we have re-parametrized the integrals over $x^+$ as integrals over $t$ along the particle trajectory $x(t)$. 
We can perform the same derivation for the other transverse coordinate $z$. The general result reads
\begin{align} \label{eq:psquared_correlator_nonabelian_app}
&\ev{p^2_i(t)}_R = \nonumber \\
&\quad\frac{2 g^2}{D_R} \intop_0^t dt' \intop_0^t dt'' \ev{\Tr \bigg[  \widetilde{F}_{i+}(x(t')) \widetilde{F}_{i+}(x(t''))\bigg]_R }.
\end{align}
\textcolor{white}{.} 
\vspace{1em}

\bibliographystyle{apsrev4-1} 
\bibliography{references} 

\end{document}